\documentclass[final,5p,times,twocolumn]{elsarticle}
\usepackage{amsmath}
\usepackage{amssymb}
\usepackage{graphicx}
\usepackage{color}
\usepackage{xspace}

\usepackage[mathscr,scaled=1.15]{urwchancal}
\DeclareFontFamily{OT1}{pzc}{}
\DeclareFontShape{OT1}{pzc}{m}{it}%
{<-> s * [1.15] pzcmi7t}{}
\DeclareMathAlphabet{\mathpzc}{OT1}{pzc}{m}{it}

\definecolor{purple}{rgb}{0.5,0,0.5}
\definecolor{blue}{rgb}{0.0,0,0.9}

\biboptions{sort&compress}

\journal{Physics Letters B}

\begin{document}

\begin{frontmatter}

\title{Natural constraints on the gluon-quark vertex}

\author[ECT]{Daniele Binosi}
\author[NU]{Lei Chang}
\author[UV]{Joannis Papavassiliou}
\author[ANL]{Si-Xue Qin}
\author[ANL]{Craig D.~Roberts}

\address[ECT]{European Centre for Theoretical Studies in Nuclear Physics
and Related Areas (ECT$^\ast$) and Fondazione Bruno Kessler\\ Villa Tambosi, Strada delle Tabarelle 286, I-38123 Villazzano (TN) Italy}
\address[NU]{School of Physics, Nankai University, Tianjin 300071, China}
\address[UV]{Department of Theoretical Physics and IFIC, University of Valencia and CSIC, E-46100, Valencia, Spain}
\address[ANL]{Physics Division, Argonne National Laboratory, Argonne, Illinois 60439, USA}

\date{18 August 2016}
%\date{11 August 2016}

\begin{abstract}
In principle, the strong-interaction sector of the Standard Model is characterised by a unique renormalisation-group-invariant (RGI) running interaction and a unique form for the dressed--gluon-quark vertex, $\Gamma_\mu$; but, whilst much has been learnt about the former, the latter is still obscure.  In order to improve this situation, we use a RGI running-interaction that reconciles both top-down and bottom-up analyses of the gauge sector in quantum chromodynamics (QCD) to compute dressed-quark gap equation solutions with 1,660,000 distinct \emph{Ans\"atze} for $\Gamma_\mu$.  Each one of the solutions is then tested for compatibility with three physical criteria and, remarkably, we find that merely 0.55\% of the solutions survive the test.  Plainly, therefore, even a small selection of observables places extremely tight bounds on the domain of realistic vertex \emph{Ans\"atze}.  This analysis and its results should prove useful in constraining insightful contemporary studies of QCD and hadronic phenomena.
\end{abstract}

\end{frontmatter}

%\tableofcontents

%\section*{Introduction}
\noindent\textbf{1.$\;$Introduction}.
The vast bulk of observable mass is generated by quantum chromodynamics (QCD), the strong-interaction sector of the Standard Model of Particle Physics (SM); and it has been known for more than fifty years that strongly-interacting theories can generate such mass dynamically \cite{Nambu:1961tp}.  Discussions of this phenomenon of dynamical chiral symmetry breaking (DCSB) %-- the generation of \emph{mass from nothing} -- 
within the context of gluons dressing the quark propagator, began with Refs.\,\cite{Lane:1974he, Politzer:1976tv} and have since continued vigorously.

A natural tool for analyses of DCSB is the gap equation for the dressed-quark Schwinger function:
\begin{subequations}
\label{gendseN}
\begin{align}
%S^{-1}(k) & = [i\gamma\cdot k + M(k^2)]/Z(k^2) \\
S^{-1}(k) & = i\gamma\cdot k \, A(k^2) + B(k^2) \\
%
%S^{-1}(p)
& = Z_2 \,(i\gamma\cdot k + m^{\rm bm}) + \Sigma(k)\,,\\
\Sigma(k)& =  Z_1 \int^\Lambda_{dq}\!\! g^2 D_{\mu\nu}(k-q)\frac{\lambda^a}{2}\gamma_\mu S(q) \Gamma^a_\nu(k,p) ,
\end{align}
\end{subequations}
where the dressed-gluon propagator may be written\footnote{
Landau gauge is typically used because it is, \emph{inter alia} \protect\cite{Bashir:2008fk,Bashir:2009fv,Raya:2013inaS}: a fixed point of the renormalisation group; that gauge for which sensitivity to model-dependent differences between \emph{Ans\"atze} for the gluon-quark vertex are least noticeable; and a covariant gauge, which is readily implemented in simulations of lattice-regularised QCD. Importantly, gauge covariance of Schwinger functions obviates any question about the gauge dependence of gauge invariant quantities.}
\begin{equation}
\label{DressedGluon}
D_{\mu\nu}(k) = \Delta(k^2) D_{\mu\nu}^{0}(k) \,,
%= \left[\delta_{\mu\nu}-\frac{k_\mu k_\nu}{k^2} \right] \frac{\Delta(k^2)}{k^2}
\end{equation}
with $D^{0}_{\mu\nu}(p)$ denoting the free-gauge-boson propagator; $\Gamma_\nu^a=(\lambda^a/2) \Gamma_\nu$ is the quark-gluon vertex; $\int^\Lambda_{dq}$, a symbol representing a Poincar\'e invariant regularisation of the four-dimensional integral, with $\Lambda$ the regularisation mass-scale; $m^{\rm bm}(\Lambda)$, the current-quark bare mass; and $Z_{1,2}(\zeta^2,\Lambda^2)$, respectively, the vertex and quark wave-function renormalisation constants, with $\zeta$ the renormalisation point, which is $\zeta_{19}=19\,$GeV herein \cite{Chang:2013pq, Maris:1997tm, Qin:2011dd, Qin:2011xq}.

Whether or not DCSB emerges in the SM is decided by the structure of the gap equation's kernel.  Hence the basic question is: Just what form does that kernel take?  Owing to asymptotic freedom, the answer is known on the perturbative domain \cite{Jain:1993qh, Maris:1997tm, Qin:2011dd, Qin:2011xq, Bloch:2002eq}, \emph{viz}.\  on ${\mathpzc A} = \{(k,q)\,|\, p^2=(k-q)^2 \simeq k^2 \simeq q^2 \gtrsim 2\,{\rm GeV}^2\}$:
\begin{align}
g^2 D_{\mu\nu}(p) \, Z_1 \, \Gamma_\nu(q,k)
& \stackrel{p^2\gtrsim 2\,{\rm GeV}^2}{=} 4\pi\alpha_s(p^2)\,
D^{0}_{\mu\nu}(p) \, Z_2^2 \,\gamma_\nu\,,
\label{UVmodelindependent}
\end{align}
where $\alpha_s(p^2)$ is QCD's running coupling.   The question thus actually relates only to the infrared domain.%, which is a complement of ${\mathpzc A}$.

Much has been learnt about the infrared behaviour of the running coupling and dressed-gluon propagator in the past two decades \cite{Aguilar:2008xm, Boucaud:2011ug, Aguilar:2015bud}; and the current state of understanding is described in Ref.\,\cite{Binosi:2014aea}.  Namely, one may write:
\begin{equation}
\label{Bridging1}
Z_1 g^2 D_{\mu\nu}(p) = Z_2 \, 4\pi \, p^2 \hat d(p^2) D^{0}_{\mu\nu}(p) \,,
\end{equation}
where $\hat d(p^2)$ is a renormalisation-group-invariant (RGI) interaction strength, which is expressed as follows:
\begin{equation}
\label{Bridging2}
p^2 \hat d(p^2) = \frac{\alpha_s(\zeta^2) \Delta(p^2;\zeta^2)}{[1+G^2(p^2;\zeta^2)]^2}\,,
\end{equation}
where $\Delta(p^2;\zeta^2)$ appears in Eq.\,\eqref{DressedGluon} and $G(p^2;\zeta^2)$ defines the transverse component of the gluon-ghost vacuum polarisation, $\Lambda_{\mu\nu}(p)$ that arises in contemporary applications of the pinch-technique (PT) \cite{Cornwall:1981zr,Cornwall:1989gv} to QCD's gauge sector \cite{Binosi:2002ez, Aguilar:2008xm, Binosi:2009qm}.

The interaction in Eq.\,\eqref{Bridging2} has been computed \cite{Aguilar:2009nf} and the primary conclusion of Ref.\,\cite{Binosi:2014aea} is that the following representation may be used with good accuracy:
\begin{equation}
\label{dhatform}
\hat d(s) \approx
\frac{2 \pi}{\omega^5} \varsigma^3 \, {\rm e}^{-s/\omega^2}
+ \frac{2 \pi \gamma_m\,{\cal F}(s)}{\ln [ \tau + (1+s/\Lambda_{\rm QCD}^2)^2]} ,
\end{equation}
where $\gamma_m^{N_f=4} = 12/25$, $\Lambda_{\rm QCD}=0.57\,$GeV (MOM renormalisation scheme); $\tau={\rm e}^2-1$\; ${\cal F}(s) = \{1 - \exp(-s/[4 m_t^2])\}/s$, $m_t=0.5\,$GeV; and $\varsigma = 0.55\,$GeV, $\omega=0.6\,$GeV \cite{Chang:2013pq}.

The pointwise form of Eq.\,\eqref{dhatform}, the values of $\varsigma$, $\omega$, and the connection with $\hat d(s)$ are not accidental.  Eqs.\,\eqref{Bridging1}--\eqref{dhatform} bridge a gap between thirty years of ``top-down'' continuum and lattice analyses of QCD's gauge sector and an equal period of ``bottom-up'' studies of its matter sector.  In doing so, they represent a long-sought unification of the top-down and bottom-up approaches to determining QCD's RGI interaction.  Critically, this advance was only made possible following an appreciation of just how important the dressed--gauge-boson--quark vertex is to DCSB, and vice versa, which had grown over many years, \emph{e.g}.\ Refs.\,\cite{Ball:1980ay, Williams:1989tv, Curtis:1990zs, Burden:1990mg, Dong:1994jr, Bashir:1994az, Munczek:1994zz, Stingl:1994nk, Bender:1996bb, Hawes:1998cw, Bhagwat:2004hn, Bhagwat:2004kj, Chang:2009zb, Kizilersu:2009kg, Chang:2010hb, Aguilar:2010cn, Qin:2013mta, Aguilar:2014lha, Kizilersu:2014ela, Mitter:2014wpa, Binosi:2016rxz}, combined with the ability to quantify its impact on the properties of hadrons \cite{Chang:2009zb, Chang:2010hb, Chang:2011ei, Sanchis-Alepuz:2015qra, Qin:2016fbu}.   It brings us to a point from which a new branch of enquiries should begin; namely, how sensitive are DCSB and hadron properties to the form of the dressed gluon-quark vertex?

\smallskip

\begin{figure}[t]
%\centerline{%
%\includegraphics[clip,width=0.35\linewidth]{Vertex.eps}}
\centerline{%
\includegraphics[clip,width=0.2\linewidth]{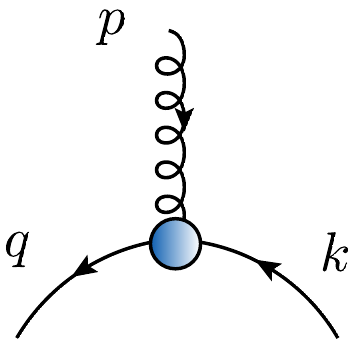}}
\caption{\label{figvertex} Gluon--quark vertex, $\Gamma_\mu(q,k)$: momentum conventions indicated.}
\end{figure}

\noindent\textbf{2.$\;$Gluon-Quark Vertex}.
The gluon-quark vertex in the SM Lagrangian is simple, \emph{viz}.\ $\Gamma_\mu^a = \Gamma_\mu^{a0} = (\lambda^a/2)\gamma_\mu$; and it remains relatively simple on ${\mathpzc A}$:  switching-on interactions, $\Gamma_\mu^{a0}$ receives momentum-dependent logarithmic corrections, but no new structures become significant.  However, the transverse Ward-Green-Takahashi (WGT) identities \cite{Takahashi:1985yz, He:2000we, Pennington:2005mw, He:2009sj, Qin:2013mta, Aguilar:2014lha} entail that the simplicity is destroyed at infrared momenta in the presence of DCSB \cite{Qin:2013mta, Aguilar:2014lha}.  In this case $\Gamma_\mu^a(q,k)$, illustrated in Fig.\,\ref{figvertex}, has up to twelve independent terms, each associated with a different Poincar\'e-covariant Dirac-matrix structure.  Half the terms in $\Gamma_\mu^a(q,k)$ are only nonzero in the chiral limit owing to DCSB, so one might expect them to act as amplifiers in the gap equation's kernel.  In such circumstances, given the role that DCSB plays in forming hadron observables, Nature must place constraints on the strength of these terms.

In order to describe one way of elucidating natural constraints on the vertex, we first decompose it $(t=q+k)$:
{\allowdisplaybreaks
\begin{subequations}
\begin{align}
\Gamma_\nu(q,k) & = \Gamma_\nu^{\rm BC} (q,k) + \Gamma_\nu^{\rm T}(q,k)\,,\\
\label{BCvertex}
i \Gamma_\nu^{\rm BC} (q,k) & = i \gamma_\nu \Sigma_A^{qk}
+ t_\nu [ i \tfrac{1}{2} \gamma\cdot t\, \Delta_A^{qk} + \Delta_B^{qk} ] =: \sum_{i=1}^3 L_\nu^i \lambda_i^{qk} \,, \\
\label{Tvertex}
\Gamma_\nu^{\rm T}(q,k) & = \sum_{i=1}^8 T_\nu^i \tau_i^{qk}\,,
\end{align}
\end{subequations}}
\hspace*{-0.5\parindent}where $\lambda_1^{qk} = \Sigma_A^{qk} = [A(q^2)+A(k^2)]/2$, $\lambda_2^{qk}= \Delta_A^{qk}$,  $\lambda_3^{qk}= \Delta_B^{qk}$,  $\Delta_F^{qk}=[F(q^2)-F(k^2)]/[q^2-k^2]$, $F=A,B$.  The first term, Eq.\,\eqref{BCvertex}, expresses that part of the vertex which satisfies the Abelian WGT identity, appropriate to the PT RGI interaction and also a fair representation of existing lattice-QCD results \cite{Bhagwat:2004kj}.

The second term, Eq.\,\eqref{Tvertex}, defined using the basis in Eq.\,\eqref{Tbasis}, describes all purely transverse contributions.  Its particular form is unknown, despite continuing efforts using continuum and lattice methods \cite{Ball:1980ay, Williams:1989tv, Curtis:1990zs, Burden:1990mg, Dong:1994jr, Bashir:1994az, Munczek:1994zz, Stingl:1994nk, Bender:1996bb, Hawes:1998cw, Bhagwat:2004hn, Bhagwat:2004kj, Chang:2009zb, Kizilersu:2009kg, Aguilar:2010cn, Qin:2013mta, Aguilar:2014lha, Mitter:2014wpa, Binosi:2016rxz, Chang:2010hb, Chang:2011ei, Skullerud:2002ge, Lin:2005zd, Kizilersu:2006et, Rojas:2013tza}; but, employing a combination of perturbation theory and the transverse WGT identities, one may construct an \emph{Ansatz} \cite{Bashir:2011dp, Qin:2013mta}:
\begin{align}
\nonumber
   \tau_1^{qk} & = a_1 \Delta_B^{qk}\,, \;
\tau_2^{qk}  = a_2 \Delta_A^{qk}\,,\;
\tau_3^{qk}  = -a_3 2 \, k\cdot q\, \Delta_A^{qk}  \,, \\
\nonumber
   \tau_4^{qk} & = a_4 \frac{4 \Delta_B^{qk}}{t^{\rm T}\cdot t^{\rm T}}\,,\;
\tau_5^{qk}  = a_5 \Delta_B^{qk} \,,\;
\tau_6^{qk}  = a_6 \frac{[q^4-k^4]\Delta_A^{qk}}{ [t\cdot p]^2 + \mu^4/4}\,, \\
%% Should be (M^2(q)+M^2(p))^2
%
   \tau_7^{qk} & = a_7 \frac{\Delta_B^{qk}}{q^2+k^2} \,,\;
 \tau_8^{qk}  = a_8 \Delta_A^{qk} \,,
\end{align}
where $\mu^2 = M^2(q^2)+M^2(k^2)$.  Here, $M(k^2) = B(k^2)/A(k^2)=:{\mathpzc m}$, the dressed-quark mass-function, is independent of the renormalisation point and $\{a_i,i=1,\ldots,8\}$ are dimensionless constants, the individual values of which modulate the strength of the associated vertex term.  This \emph{Ansatz} is simple, involving only those functions that appear in the quark propagator, but sufficiently general to allow for a meaningful exploration of the vertex' impact on the gap equation's solution and hence the constraints that observables impose thereupon.

\smallskip

\noindent\textbf{3.$\;$Natural Constraints on the Vertex}.
We are about to undertake a challenging task, so it is sensible to first remark that extant studies of the meson spectrum  \cite{Chang:2011ei, Qin:2016fbu} suggest that the $T_\nu^{2,6,7}$ terms in $\Gamma_\nu^{\rm T}$ are of lesser importance.  Hence, we neglect them hereafter, setting $a_2=a_6=a_7=0$.\footnote{As explained elsewhere \cite{Curtis:1990zs, Bashir:2011dp}, consistency with both perturbation theory and multiplicative renormalisability in QED is preserved so long as a certain additive combination of $a_3$ and $a_6$ is nonzero, so the choice $a_6=0$ is compatible with earlier studies.}  Additionally, following some straightforward algebra it becomes evident that the gap equation's kernel does not depend separately on $a_{4}$, $a_{5}$, but, instead, only on the combination $a_{\hat{45}} = a_4-3a_5$.  Allowing for these simplifications, we proceeded to scan the vertex parameter space spanned by the constants $a_{1,3,\hat{45},8}$, chosen within
\begin{align}
\nonumber
\mathbb{V}_4 & = \{(a_1,a_3,a_{\hat{45}},a_8) \\
& |\, a_1,a_3 \in [-1,1], a_{\hat{45}} \in [-7,5], a_8\in [-5,1]\}\,.
\label{vertexcoefficients}
\end{align}
As will subsequently become apparent, it is unnecessary to explore a larger subset of $\mathbb{R}^4$.

The scanning method is simple.  Working in the chiral limit,
we randomly generated a quadruplet ${\mathpzc q} =(a_1,a_3,a_{\hat{45}},a_8)$ and therewith formed the gluon-quark vertex, $\,^{\mathpzc q} \Gamma_\nu$;
solved the gap equation with that vertex and the RGI interaction in Eq.\,\eqref{dhatform};
and categorised the solutions as acceptable if:
(\emph{i}) they expressed DCSB of sufficient and reasonable strength, which means $M(0) \in (0.25,0.45)\,$GeV,
(\emph{ii}) the associated dressed-quark anomalous chromomagnetic moment (ACM) was negative-definite,\footnote{Given a monotonically decreasing dressed-quark mass function, this feature is required in order satisfy constraints of perturbative QCD (pQCD) \cite{Chang:2010hb}.}
and
(\emph{iii}) the pion's leptonic decay constant was within 5\% of its chiral-limit value, $f_\pi^0 \approx 0.088\,$GeV \cite{Gasser:1983yg}.

\begin{figure}[t]

\centerline{\includegraphics[width=0.75\linewidth]{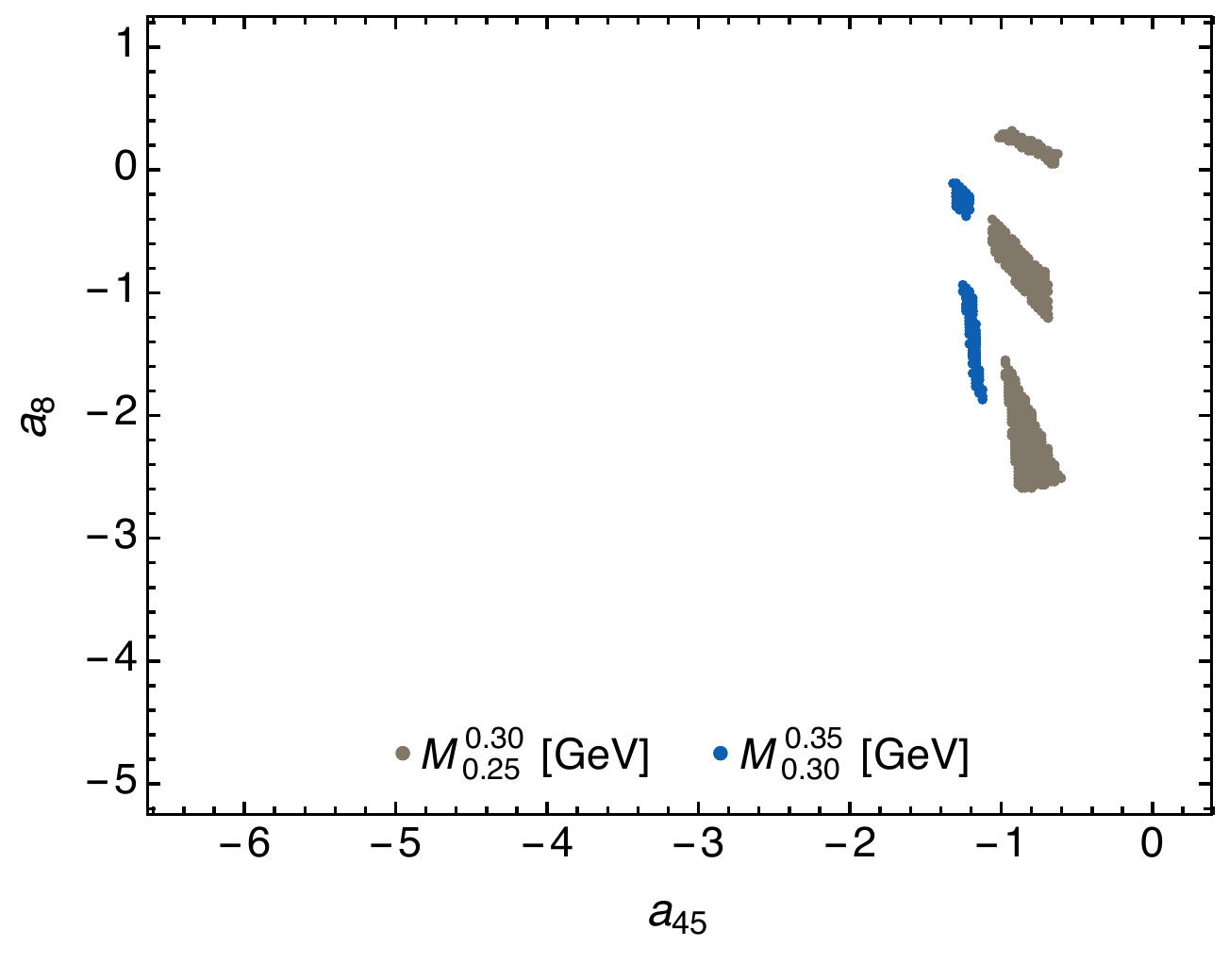}}

\vspace*{1ex}

\centerline{\includegraphics[width=0.75\linewidth]{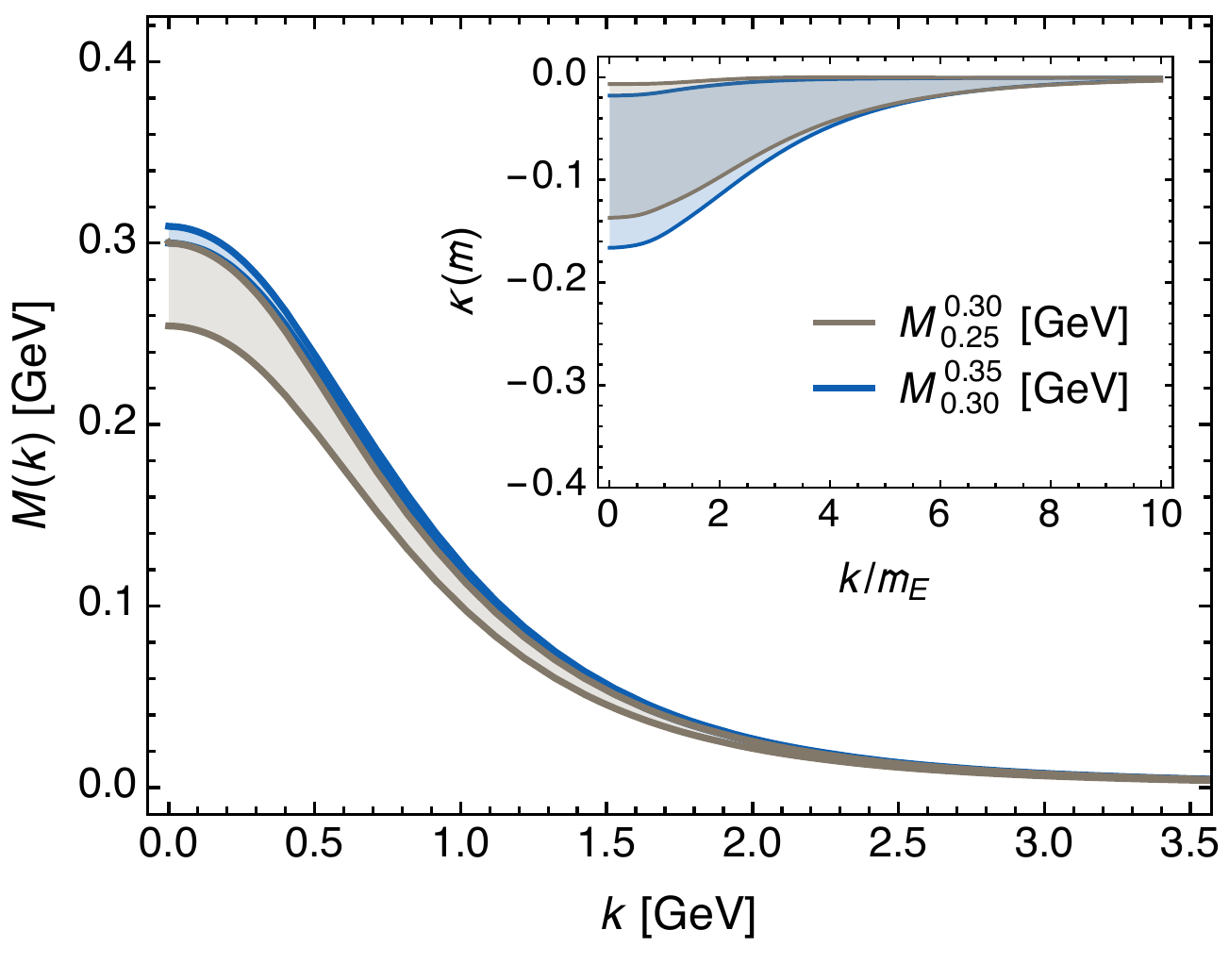}}

\caption{Case A.  \emph{Upper panel} -- Domain of coefficient values ${\mathpzc q} =(0,0,a_{\hat{45}},a_8)$ that simultaneously satisfy cuts (\emph{i})--(\emph{iii}).
The two bands indicate: light (grey), $M(0) \in (0.25,0.30)\,$GeV; dark (blue), $M(0) \in (0.30,0.35)\,$GeV.
%The four bands indicate: lightest (brown), $M(0) \in (0.25,0.30)\,$GeV; darker (red), $M(0) \in (0.30,0.35)\,$GeV; darker still (blue), $M(0) \in (0.35,0.40)\,$GeV; and darkest (navy), $M(0) \in (0.40,0.45)\,$GeV.
%
\emph{Lower panel} -- The mass functions produced by the vertex \emph{Ans\"atze} identified in the upper panel lie within the like-shaded bands; and the inset depicts the associated anomalous chromomagnetic moment (ACM), Eq.\,\eqref{ACMdistribution}.  (${\mathpzc m}_E$ is the Euclidean constituent-quark mass, \emph{i.e}.\ the solution of $M({\mathpzc m}_E) = {\mathpzc m}_E$.)
\label{CaseA}}
\end{figure}

The ACM distribution is explained in Refs.\,\cite{Chang:2010hb, Bashir:2011dp} and may be estimated using
\begin{equation}
\label{ACMdistribution}
\kappa({\mathpzc m}) =2{\mathpzc m}
\frac{[a_5-1+a_1/2] \delta_B + {\mathpzc m} (1-a_8)\delta_A}
{\sigma_A+2 {\mathpzc m}^2 (a_3-1) \delta_A +2 {\mathpzc m}(1-a_1/2)\delta_B}
\end{equation}
where $\sigma_A=\Sigma_A^{{\mathpzc m}{\mathpzc m}}$, $\delta_{A,B}=\Delta_{A,B}^{{\mathpzc m}{\mathpzc m}}$. Concerning $f_\pi$, absent a solution to the pion's Bethe-Salpeter equation, two approximating formulae exist: $f_\pi^{\rm PS}$ (Eq.\,(11), Ref.\,\cite{Pagels:1979hd}) and $f_\pi^{\rm CR}$ (Eq.\,(4.5), Ref.\,\cite{Cahill:1985mh}).  We use the latter because it is more accurate \cite{Chang:2009zb}, but the source of the differences between them is understood and they are identical in the chiral limit when the necessary corrections to both are included \cite{Maris:1997hd}.

\begin{figure}[t]

\centerline{\includegraphics[width=0.75\linewidth]{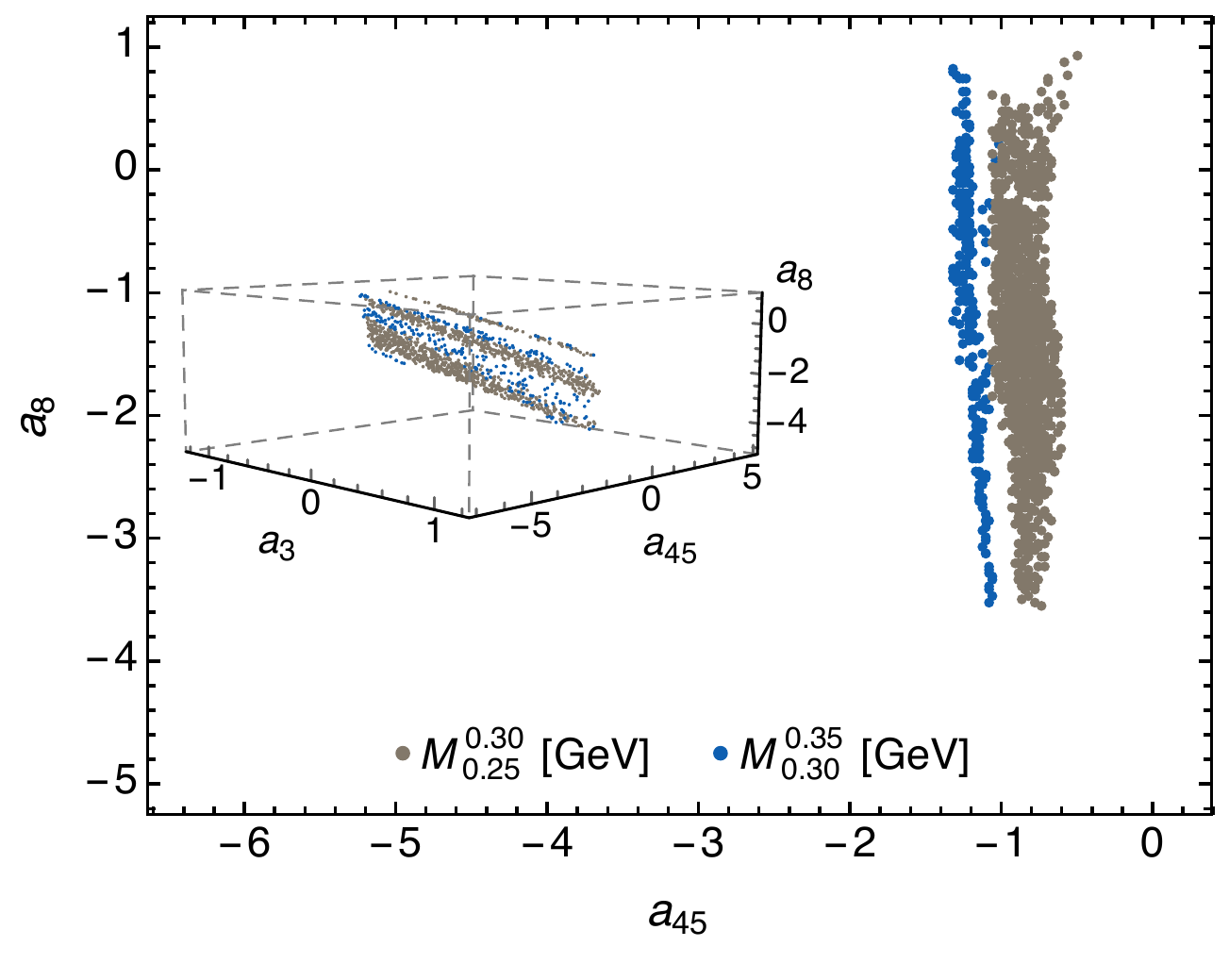}}

\vspace*{1ex}

\centerline{\includegraphics[width=0.75\linewidth]{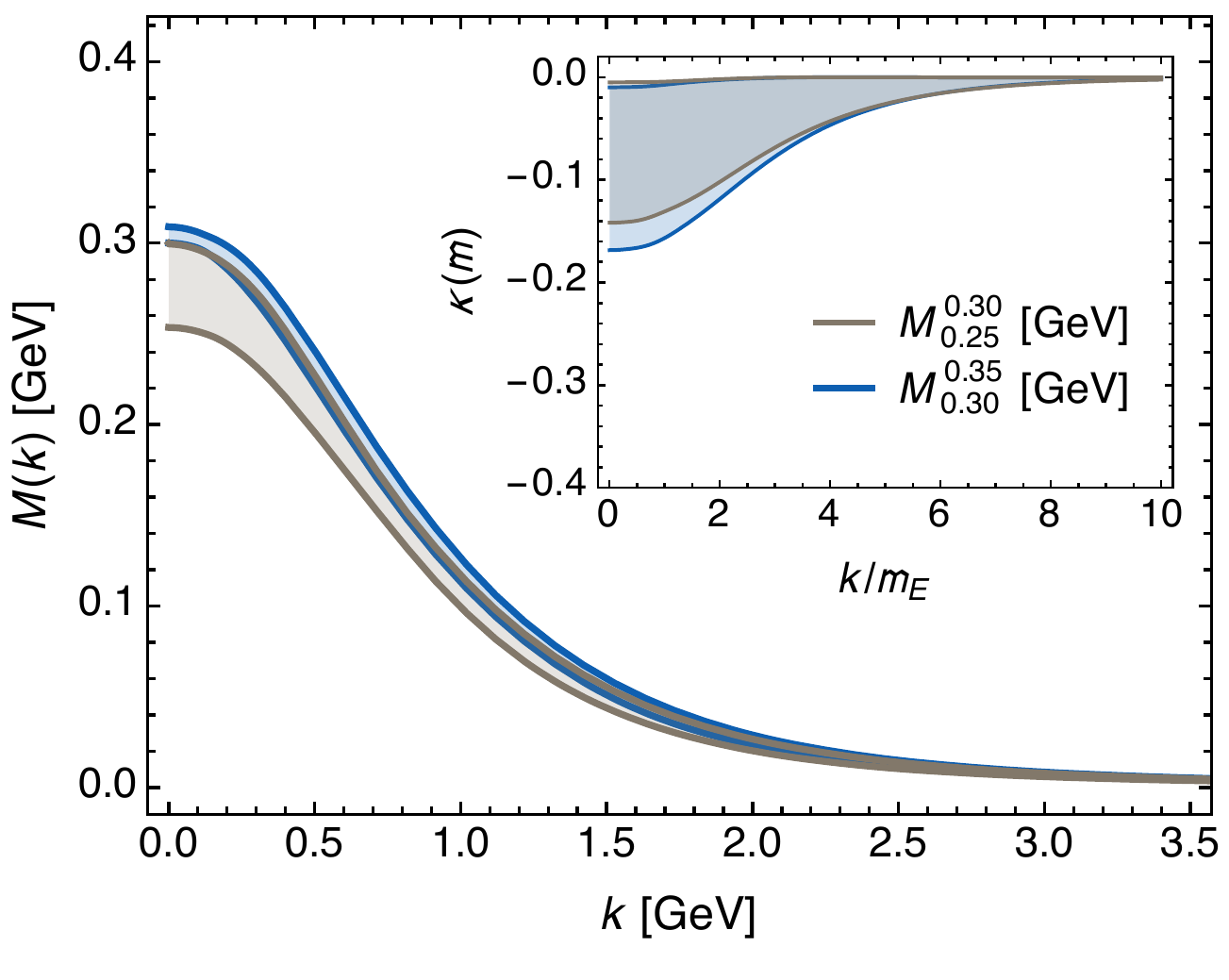}}

\caption{Case B.  \emph{Upper panel} -- Domain of coefficient values ${\mathpzc q} =(0,a_3,a_{\hat{45}},a_8)$ that simultaneously satisfy cuts  (\emph{i})--(\emph{iii}).  With the inclusion of $a_3$, the space of \emph{Ans\"atze} is three-dimensional; and the inset displays the volume that survives all cuts and the correlation between coefficients.
\emph{Lower panel} -- The mass functions generated by the vertex \emph{Ans\"atze} identified in the upper panel lie within the like-shaded bands and produce the ACM drawn in the inserted panel.
(The legend is the same as that in Fig.\,\ref{CaseA}.)
\label{CaseB}}
\end{figure}

\smallskip

\noindent{\emph{3.1}$\;$\emph{Case A}: $a_1=0=a_3$}.
Owing to the known importance of $a_{5,8}$ in determining the dressed-quark ACM \cite{Chang:2010hb}, nonzero values for these coefficients are always included in our sample space.  Therefore, we first selected 360,000 independent values of ${\mathpzc q} =(0,0,a_{\hat{45}},a_8)$, solved the gap equation in each case, and catalogued the solutions as described above.  Only 6\% of the vertex parameter space survived the first cut, (\emph{i}).  The volume fell to 3\% when (\emph{i}) and (\emph{ii}) were applied together.  Finally, just 0.4\% of the 360,000 vertex \emph{Ans\"atze} considered could simultaneously satisfy (\emph{i})--(\emph{iii}).  The paucity of such \emph{Ans\"atze} is highlighted by the upper panel of Fig.\,\ref{CaseA}: the sample space is actually larger than the planar area drawn.   The lower panel of Fig.\,\ref{CaseA} shows that in this case the three criteria we've applied enforce $M(0)\in (0.26,0.31)\,$GeV.
%%the percentage of solutions surviving the three cuts is 0.4% --rather than the current 0.5%— and M(0)\in(0.26,0.31) —rather than the current M(0)\in(0.26,0.32) .

\smallskip

\noindent{\emph{3.2}$\;$\emph{Case B}: $a_1=0$}.
Here we generated 400,000 vertex \emph{Ans\"atze}.  Proceeding as above, one obtains the domain of acceptable vertices depicted in the upper panel of Fig.\,\ref{CaseB}.
%Evidently, absent the positive feedback provided by $\tau_1^{qk}$, the mass functions generated by this class of vertices are smaller in magnitude at infrared momenta than those in Case~B,
The mass functions obtained in this instance are similar to those obtained with Case~A \emph{Ans\"atze}, as illustrated by the lower panel of Fig.\,\ref{CaseB}: the largest attainable value in solutions consistent with our criteria is $M(0)=0.31\,$GeV.  Notably, including $a_3\neq 0$, acceptable solutions can be obtained within a larger domain of $a_8$ values: these two tensor structures mutually compensate.  However, comparing the lower panels of Figs.\,\ref{CaseA} and \ref{CaseB}, one sees that $a_3\neq 0$ does not much affect the mass-function's $k^2$-dependence.

\smallskip

\begin{figure}[t]

\centerline{\includegraphics[width=0.75\linewidth]{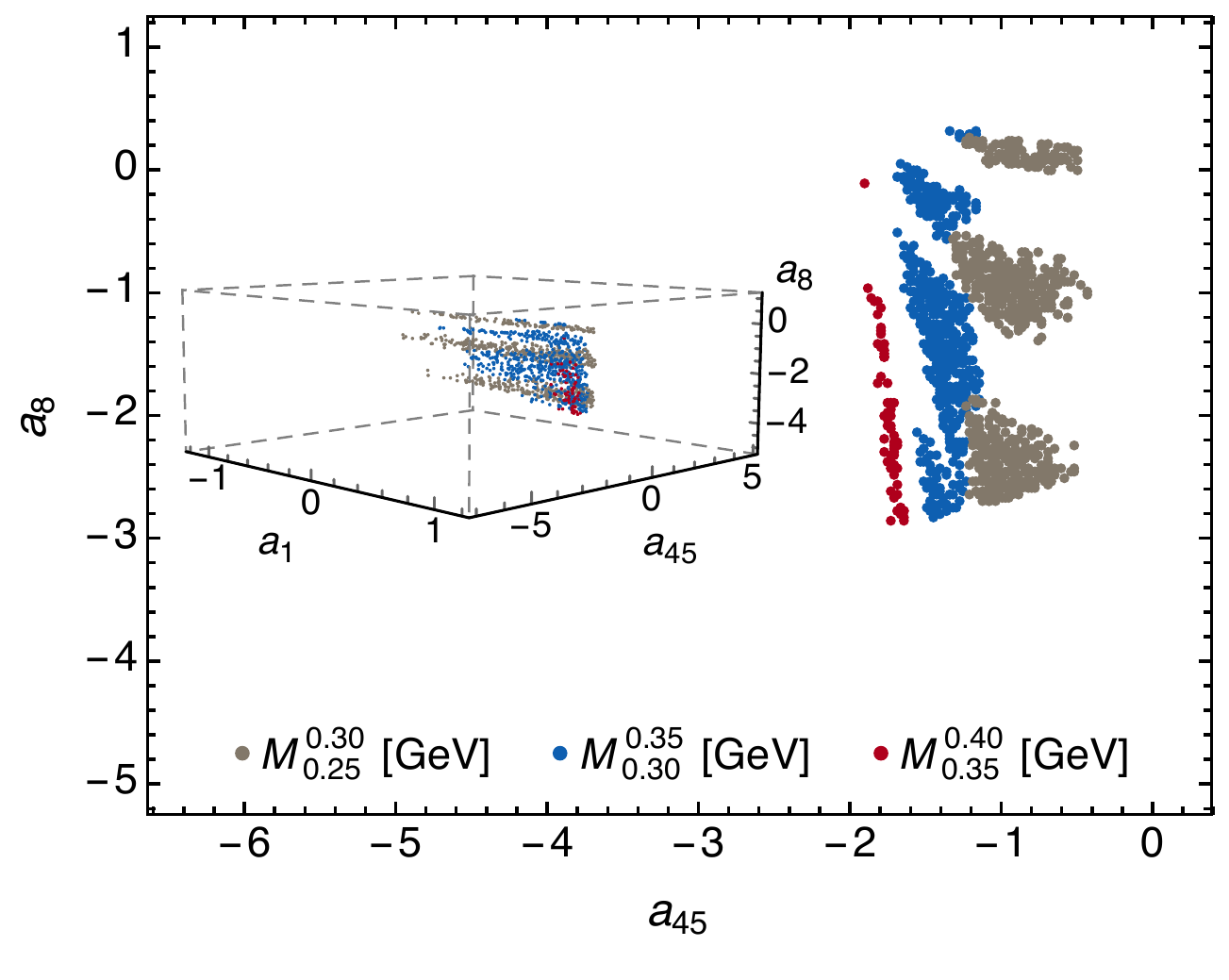}}

\vspace*{1ex}

\centerline{\includegraphics[width=0.75\linewidth]{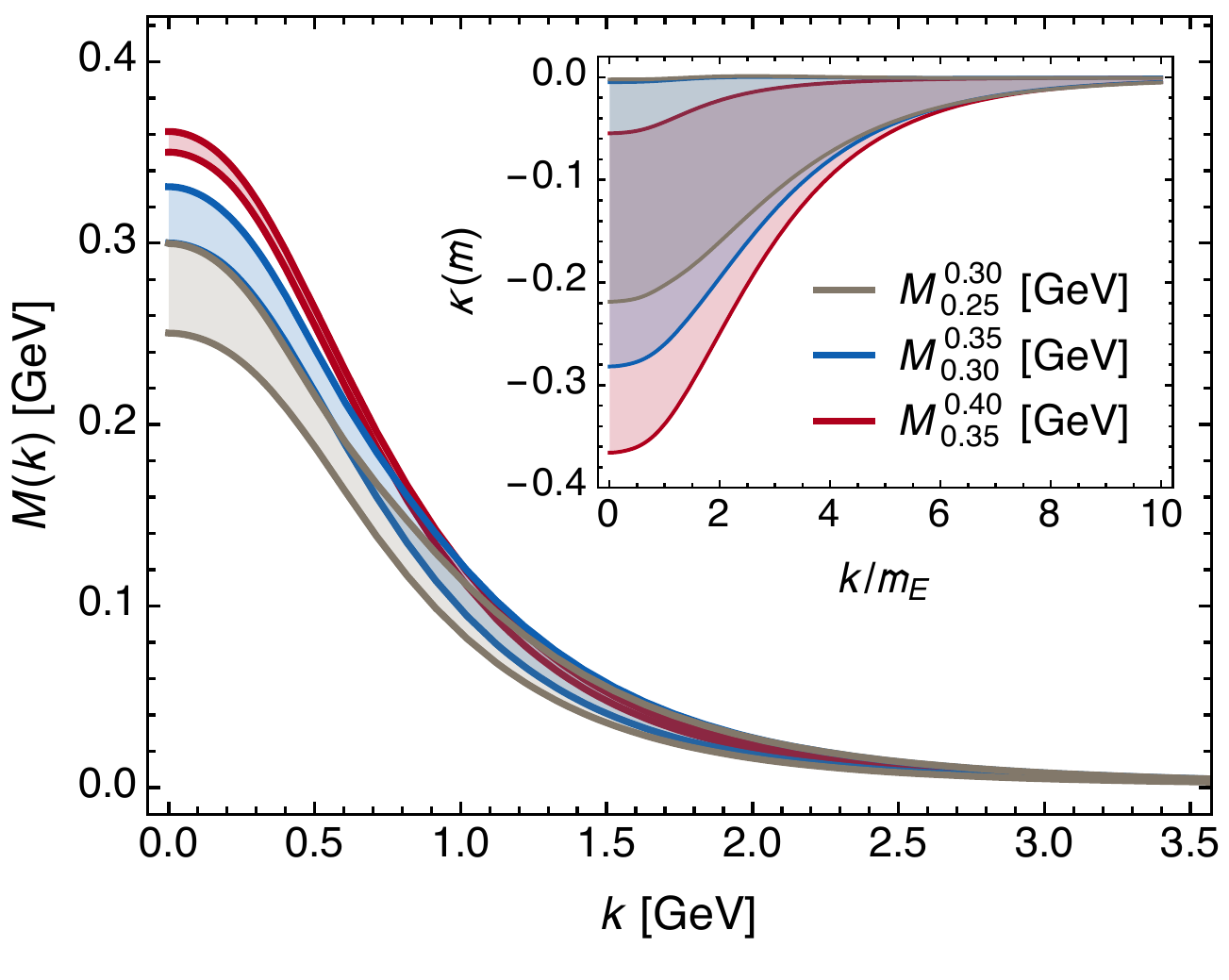}}

\caption{Case C.  \emph{Upper panel} -- Domain of coefficients values ${\mathpzc q} =(a_1,0,a_{\hat{45}},a_8)$ that simultaneously satisfy cuts  (\emph{i})--(\emph{iii}).  The space of \emph{Ans\"atze} is three-dimensional; and the inset displays the volume that survives all cuts and the correlations between coefficients.  The three bands indicate: light (grey), $M(0) \in (0.25,0.30)\,$GeV; dark (blue), $M(0) \in (0.30,0.35)\,$GeV; darker (red), $M(0) \in (0.35,0.40)\,$GeV.  Very few acceptable vertices lie in the last category, denoted by the leftmost points.
\emph{Lower panel} -- The mass functions generated by the vertex \emph{Ans\"atze} identified in the upper panel lie within the like-shaded bands and produce the ACM drawn in the inserted panel.
%The inset illustrates the effect of $a_1$: for a fixed value of $M(0)$, a smaller value of $a_1$ (solid curve) yields a mass-function that runs more slowly with $k$ than does a larger value of $a_1$ (dashed curve).
%
\label{CaseC}}
\end{figure}

\noindent{\emph{3.3}$\;$\emph{Case C}: $a_3=0$}.
In this instance we selected 360,000 independent values of ${\mathpzc q} =(a_1,0,a_{\hat{45}},a_8)$ and proceeded as before, with the result expressed in the upper panel of Fig.\,\ref{CaseC}.  Qualitatively, the outcome is similar to Case~A.  Successive cuts progressively restrict the space of acceptable vertex \emph{Ans\"atze}, so that finally $M(0)\in (0.25,0.36)\,$GeV.

A new feature is expressed in the effect of $a_1$ on $M(k^2)$: for a fixed value of $M(0)$, an increase in $a_1$ produces a steeper decline in $M(k^2)$.  [We return to this point in connection with Fig.\,\ref{a1Massk2} below.]
This might have been anticipated because, with $a_1\neq 0$, the vertex includes a new term that depends linearly on both the magnitude and $k^2$-dependence of the dynamically generated mass function.  It follows \cite{Cloet:2013gva} that the associated vertex contribution can influence whether or not the proton's electric form factor possesses a zero and, conversely, can be constrained by the pointwise behaviour of that form factor.

\smallskip

\begin{figure}[t]

\centerline{\includegraphics[width=0.75\linewidth]{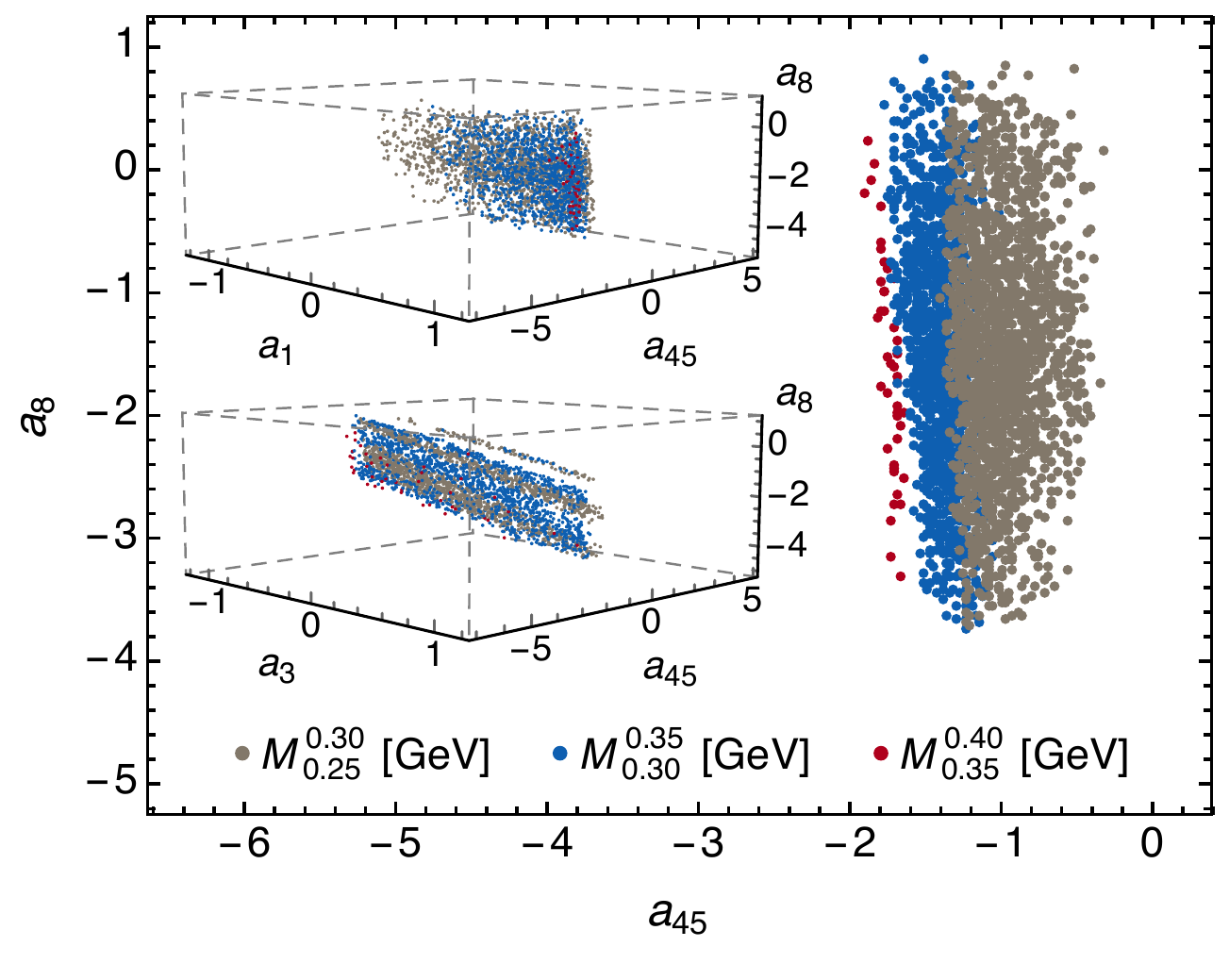}}

\vspace*{1ex}

\centerline{\includegraphics[width=0.75\linewidth]{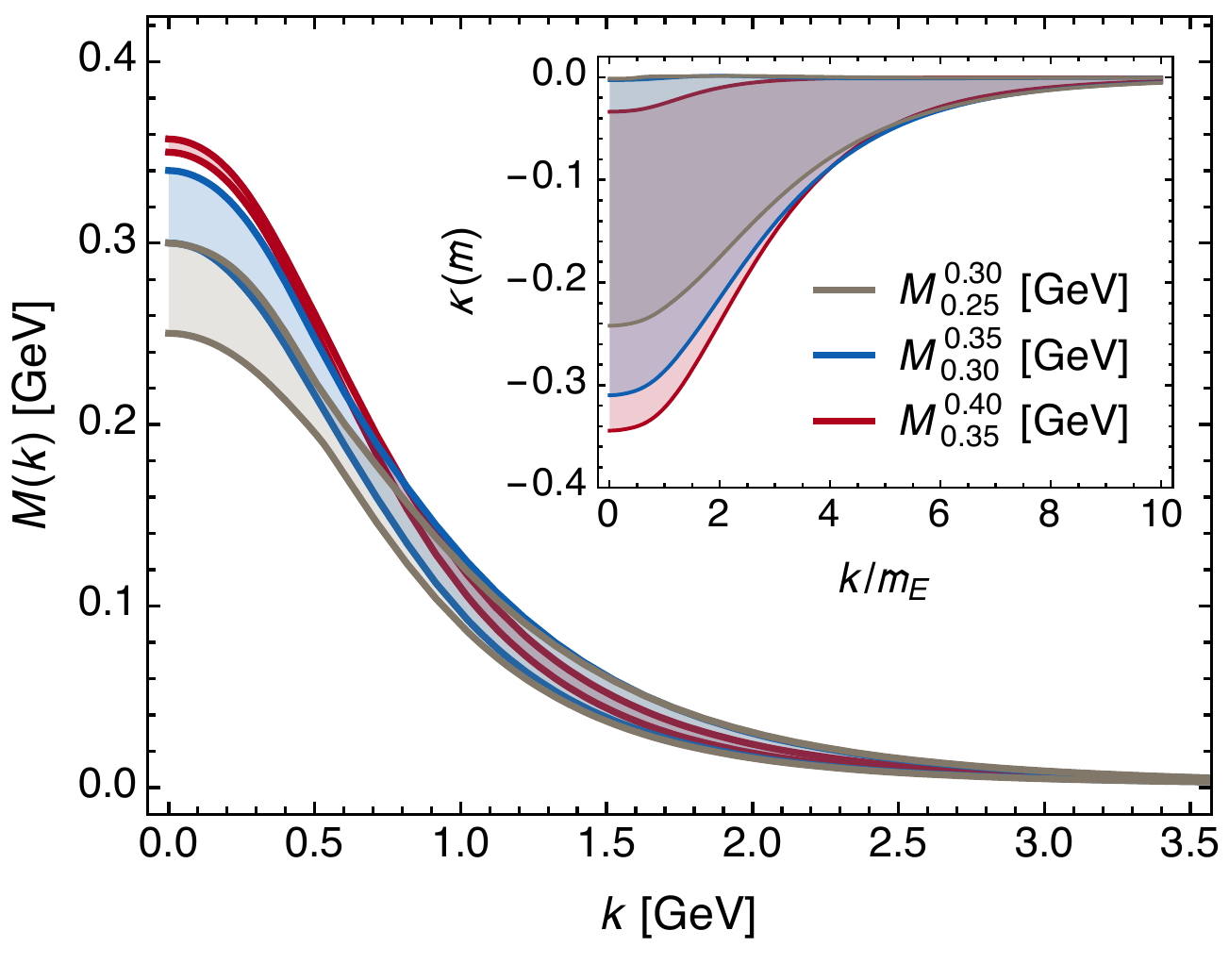}}

\caption{Case D.  \emph{Upper panel} -- Domain of parameter values that simultaneously satisfy cuts  (\emph{i})--(\emph{iii}).  With all coefficients active, the space of \emph{Ans\"atze} is four-dimensional; and the two insets assist in displaying the hyper-volume that survives all cuts and the correlations between the coefficients.
\emph{Lower panel} -- The mass functions produced by the vertex \emph{Ans\"atze} identified in the upper panel lie within the like-shaded bands.
(The legend is the same as that in Fig.\,\ref{CaseB}.)
\label{CaseD}}
\end{figure}

\noindent{\emph{3.4}$\;$\emph{Case D -- all four coefficients nonzero}}.
With all coefficients in Eq.\,\eqref{vertexcoefficients} nonzero, we randomly selected 540,000 vertices, solved the gap equation in each case, and filtered the solutions via criteria (\emph{i})--(\emph{iii}).  Only 5.6\% of the parameter space survived the first cut, (\emph{i}); the fraction fell to 2.8\% when (\emph{i}) and (\emph{ii}) were applied together; and just 0.55\% of the 540,000 vertex \emph{Ans\"atze} considered could simultaneously satisfy (\emph{i})--(\emph{iii}).  The allowed \emph{Ans\"atze} are identified in the upper panel of Fig.\,\ref{CaseD}, with the associated mass functions depicted in the lower panel.

All features highlighted already are preserved: changes in $a_3$ and $a_8$ compensate each other, but otherwise $a_3$ is only loosely constrained by the criteria; $a_1$, $a_{\hat{45}}$, $a_8$ influence the value of observables and are thereby tightly constrained; and, within the space we have searched, only $a_1$ has a material impact on the $k^2$-dependence of the mass function, which is illustrated in Fig.\,\ref{a1Massk2}.

So as to further highlight the connection we drew in Sec.\,3.3 between the action of $a_1$ and a zero in the proton's electric form factor, $G_{Ep}$, Fig.\,\ref{a1Massk2} also depicts a comparison between the impact of $a_1$ on the $k^2$-dependence of the mass-function (solid curve) and that \emph{assumed} for the effect of DCSB and vertex feedback in Ref.\,\cite{Cloet:2013gva} (dashed curves).   The similarity between the curves indicates that the conclusions drawn in Ref.\,\cite{Cloet:2013gva} are sound; namely, that the possible existence and location of a zero in $G_{Ep}$  are
a fairly direct measure of the structure and momentum-dependence of the quark-quark scattering kernel.
%alpha=1 cf. 1.4  & alpha=1 cf. 2

It is also worth remarking that the magnitude of $M(0)$ is largely determined by $a_{\hat{45}}$: given a value of this coefficient, the other three can vary within material subdomains of their search space without measurably affecting $M(0)$.  For instance, a value of $M(0)=0.275\,$GeV is maintained to within 0.2\% when $a_{\hat{45}}\in [-1.16,-0.68]$ (4\%) and $a_1\in [-0.6,1.0]$ (80\%), $a_3\in [-1,1]$ (100\%), $a_8\in [-3.4,0.3]$ (60\%), where the parenthesised percentages indicate the size of the subdomain relative to that specified for the coefficient in Eq.\,\eqref{vertexcoefficients}.
%%-1.23774 +- 0.0661137

The behaviour of the dressed-quark wave function renormalisation, $Z(p^2) = 1/A(p^2)$, is often overlooked in studies of DCSB and the gluon-quark vertex, possibly because in many analyses some of its effects are (implicitly) absorbed into the model interaction.  This can lead, however, to results for $Z(p^2)$ that conflict with known constraints, \emph{e.g}.\ pQCD requires that in Landau gauge $Z(p^2)\to 1^-$  as $p^2\to {\zeta^2}^-$ \cite{Davydychev:2000rt}.  This feature is preserved by all \emph{Ans\"atze} that survive the cuts we apply.  It is not sufficient, however, to ensure $Z(p^2)$ is monotonic on $p^2\in[0,\zeta^2]$.  That outcome is only guaranteed if one also requires $M(0) \gtrsim 0.35\,$GeV.

\begin{figure}[t]

\centerline{\includegraphics[width=0.75\linewidth]{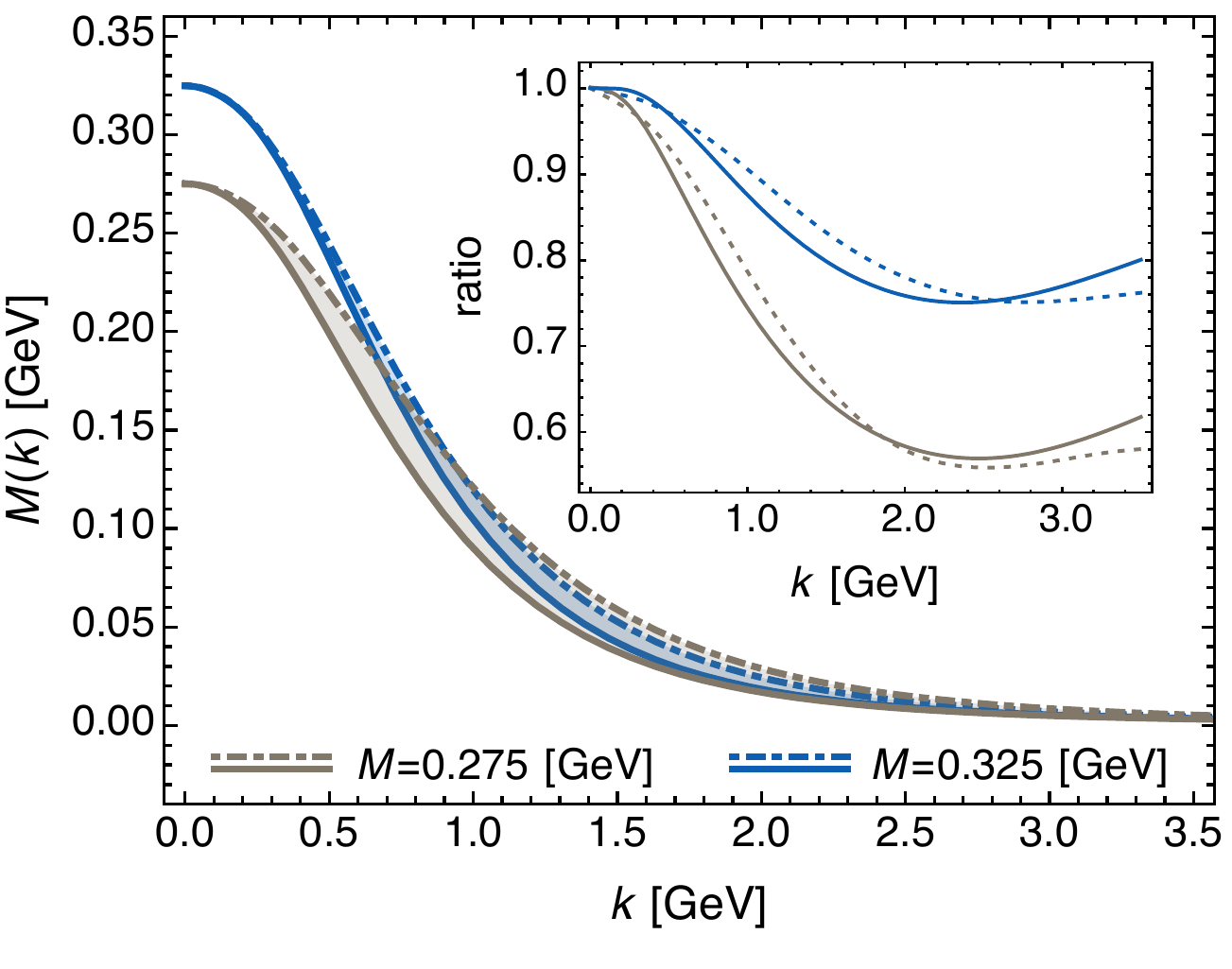}}

\caption{Impact of $a_1$ on the behaviour of $M(k^2)$.  For a fixed value of $M(0)$, a larger value of $a_1$ yields a mass-function that runs more rapidly with $k$ ($M_f(k)$, solid curves) than does a smaller value of $a_1$ ($M_s(k)$, dot-dashed curves).
\emph{Inset}: Comparison of this effect, as exhibited in ratio$\,=M_f(k)/M_s(k)$, with that assumed in Ref.\,\cite{Cloet:2013gva} --
solid light (grey) curves depict ratio computed from the like curves in the main figure, dashed light curve is ratio as obtained in Ref.\,\cite{Cloet:2013gva} using $\alpha=2$, the largest suppression of DCSB considered therein.  ($\alpha=1$ means no suppression.);
solid dark (blue) curves, obtained from like curves in the main figure, and dark dashed curve, ratio associated with $\alpha=1.4$ in Ref.\,\cite{Cloet:2013gva}.
\label{a1Massk2}}
\end{figure}

\begin{figure}[t]

\centerline{\includegraphics[width=0.75\linewidth]{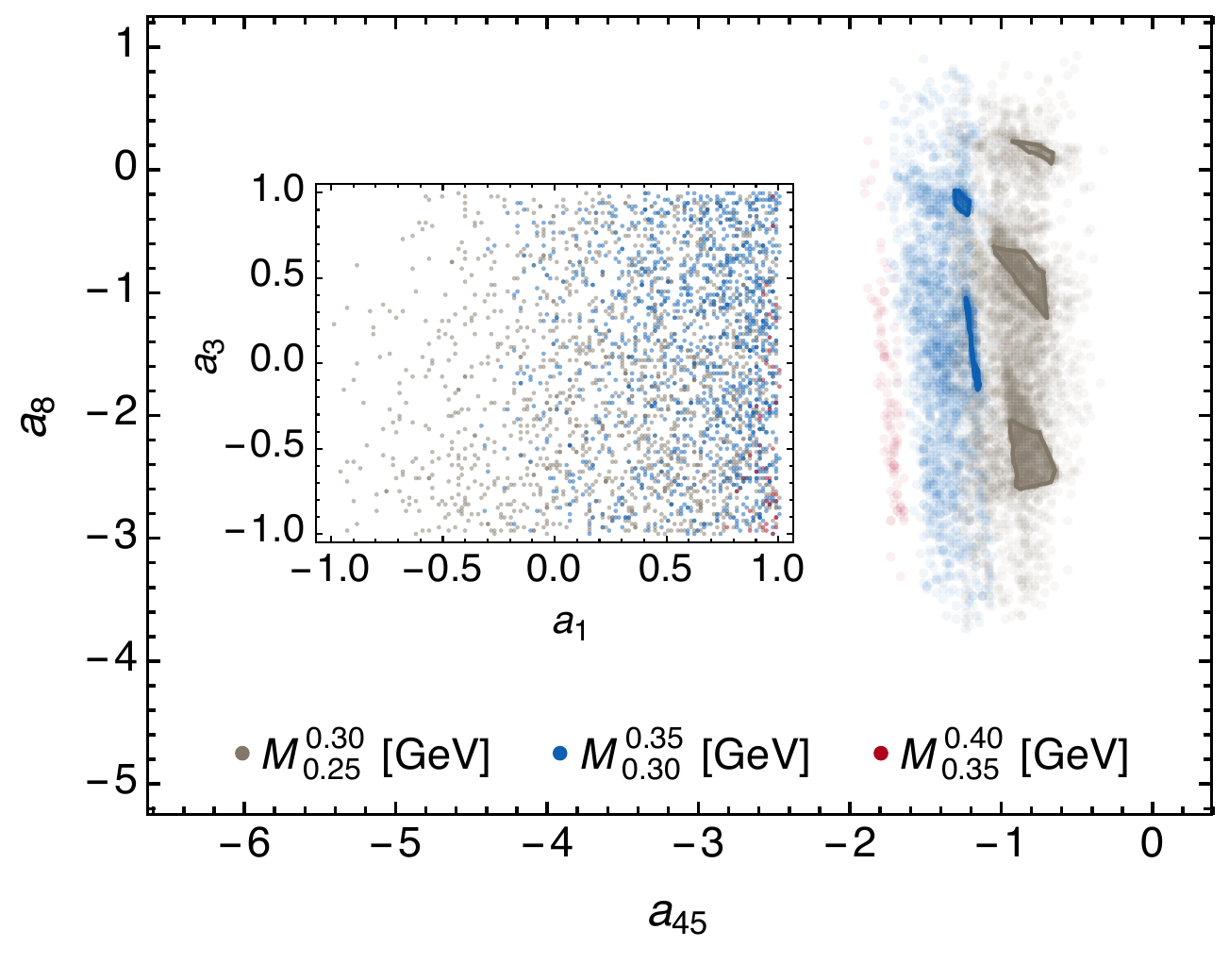}}

\caption{
Combination of all domains of acceptable vertex coefficient values depicted in Figs.\,\ref{CaseA}--\ref{CaseD}.   The regions of highest intensity identify neighbourhoods of similar (not identical) vertex coefficient values explored in all four cases, whereas the lowest intensity regions indicate domains reached only in studying Case~D.
\emph{Inset} -- Distribution of allowed vertices in the $(a_1,a_3)$ plane.  Evidently, criteria (\emph{i})--(\emph{iii}) place little constraint on $a_3$.  On the other hand, the value of $a_1$ has a strong influence on the feedback necessary for DCSB, with larger values of $a_1$ yielding a mass function that is larger at $k^2=0$, but also falls more quickly towards its ultraviolet limit as $k^2$ increases (Fig.\,\ref{a1Massk2}).
Legend as in Figs.\,\ref{CaseC}, \ref{CaseD}, \emph{viz}.\ light (grey), $M(0) \in (0.25,0.30)\,$GeV; dark (blue), $M(0) \in (0.30,0.35)\,$GeV; darker (red), $M(0) \in (0.35,0.40)\,$GeV.
\label{Casea1a3}}
\end{figure}

In Fig.\,\ref{Casea1a3} we depict both the combined results from Cases~A--D, with neighbourhoods of similar vertex coefficient values ${\mathpzc q}$ sampled in all cases indicated by the domains of highest intensity, and highlight the interplay between our filtering criteria and the coefficients $a_{1,3}$.  Plainly, the criteria do not place tight dynamical constraints on $a_3$, although, as remarked earlier, the presence of $a_3 \neq 0$ does permit a larger domain of $a_8$ values in acceptable \emph{Ans\"atze}.  In connection with $a_1$, on the other hand, the allowed range of values depends on the magnitude of $M(0)$: $-0.5 \lesssim a_1 \lesssim  1$ for $M(0) \in (0.26,0.36)\,$GeV.

If one now denotes by $\mathbb{G}_4$ that subdomain of vertex \emph{Ans\"atze}, in the class we have considered, whose members each yield a gap equation solution consistent with criteria (\emph{i})--(\emph{iii}), then
\begin{align}
\nonumber
\mathbb{G}_4 & \subset \{(a_1,a_3,a_{\hat{45}},a_8) \, | \,
    a_1\in [-0.5,1], \\
&  a_3\in [-1,1], a_{\hat{45}}\in [-2,-0.4]\,, a_8\in [-4,1]\} \subset \mathbb{V}_4\,.
\end{align}

It is obvious but nevertheless worth highlighting that the bare vertex, $\Gamma_\mu^{a0}$, is not a member of $\mathbb{G}_4$.  In fact, using the RGI running-interaction explained in the Introduction, which unifies the top-down and bottom-up approaches to charting QCD's gauge sector, the bare vertex is incapable of inducing DCSB.  Positive feedback, generated by, \emph{e.g}.\ the $L_\nu^3$ and $T_{4,5}$ terms in $\Gamma_\mu^a$, is necessary to achieve DCSB with a realistic interaction.
Plainly, therefore, in order to secure a successful description of some subset of hadron observables with the widely-used rainbow-ladder DSE truncation \cite{Munczek:1994zz, Bender:1996bb, Maris:1997tm, Maris:1999nt}, the gap equation's kernel must include an unrealistic magnification at infrared-momenta.  Hence, one should be cautious when developing an interpretation of results obtained in this way, \emph{e.g}.\ those related directly to the pointwise behaviour of bound-state Bethe-Salpeter amplitudes will, at most, only be semi-quantitatively reliable \cite{Nguyen:2010yh, Chang:2013pq, Cloet:2013tta, Ding:2015rkn, Horn:2016rip}, and that will also impact upon level-ordering and -spacing in the hadron spectrum \cite{Chang:2009zb, Krassnigg:2009zh, Krassnigg:2010mh, Roberts:2011cf, Chang:2011ei, Qin:2011dd, Qin:2011xq, Chen:2012qr, Rojas:2014aka, Gomez-Rocha:2015qga, Eichmann:2016yit}.

\smallskip

\noindent\textbf{4.$\;$Epilogue}.
Using a renormalisation-group-invariant (RGI) running-interaction that reconciles both top-down and bottom-up analyses of the gauge sector in quantum chromodynamics (QCD), we computed dressed-quark gap equation solutions with 1.66-million distinct \emph{Ans\"atze} for the dressed--gluon-quark vertex, $\Gamma_\mu$.  The \emph{Ans\"atze} were selected from a class whose members can uniquely be identified by a vector in $\mathbb{R}^4$, and those we studied were selected at random from a compact subdomain, $\mathbb{V}_4$, of this space whose limits were chosen deliberately in order to ensure consistency with extant explorations of the gauge-boson--fermion three-point function in quantum field theory.

Each member of the set of gap equation solutions thus obtained was tested for its compatibility with three physical criteria: (\emph{i}) does it express a sufficient and reasonable amount of dynamical chiral symmetry breaking (DCSB); (\emph{ii}) is the associated dressed-quark anomalous chromomagnetic moment negative-definite; and (\emph{iii}) does it produce a value for the pion's leptonic decay constant that lies within 5\% of its chiral-limit value?  Remarkably, merely 0.55\% of the solutions survived the test.
One may therefore conclude that even a small selection of observables places very tight constraints on the domain of realistic vertex \emph{Ans\"atze}, $\mathbb{G}_4$, so that $\mu(\mathbb{G}_4)\approx 0$ within ${\mathbb R}^4$, \emph{i.e}.\ the hyper-volume occupied by the space of physically acceptable vertices is extremely small.

Of course, within the Standard Model there is a unique RGI running-interaction and a unique form for $\Gamma_\mu$.  However, so long as it is necessary for studies of hadron properties to employ assumptions about the gluon-quark vertex, then our results will help to ensure those assumptions are realistic.

%%%%%%%%%%%%%%%%%%%%
%%   Acknowledgments   %%
%%%%%%%%%%%%%%%%%%%%
\smallskip

\noindent\textbf{Acknowledgments}.
DB acknowledges correspondence with H.\,Sanchis-Alepuz and D.\,Gazda.
The results described herein were obtained using the KORE HPC of the Fondazione Bruno Kessler.
This research was supported by:
Spanish MEYC under grants FPA2014-53631-C2-1-P and SEV-2014-0398, and Generalitat Valenciana  under grant Prometeo~II/2014/066;
Argonne National Laboratory, Office of the Director, through the Named Postdoctoral Fellowship Program;
and U.S.\ Department of Energy, Office of Science, Office of Nuclear Physics, contract no.~DE-AC02-06CH11357.

\smallskip

\setcounter{equation}{0}
\renewcommand{\theequation}{A\arabic{equation}}

\noindent\textbf{Appendix}.
Here we list the tensors used in Eq\,\eqref{Tvertex}:
\begin{align}
\nonumber
   T_\nu^1 & = \tfrac{i}{2} t_\nu^{\rm T}\,, \;
   T_\nu^2  = -i T_\nu^1 \gamma\cdot t\,, \;
   T_\nu^3   = \gamma_\nu^{\rm T} \,, \\
\nonumber
   T_\nu^4 & = -i T_\nu^1 \sigma_{\alpha\beta} q_\alpha k_\beta\,,\;
   T_\nu^5    = \sigma_{\nu\rho} p_\rho \,,\;
   T_\nu^6    = t_\nu \gamma\cdot p - \gamma_\nu t\cdot p\,,\\
\nonumber
T_\nu^7 & = \tfrac{i}{2} t\cdot p [\gamma_\nu \gamma\cdot t - t_\nu] - t_\nu \sigma_{\alpha\beta} q_\alpha k_\beta \,,\\
T_\nu^8 & = q_\nu \gamma\cdot k - k_\nu \gamma\cdot q + i \gamma_\nu \sigma_{\alpha\beta} q_\alpha k_\beta\,,
\label{Tbasis}
\end{align}
%where $ p\cdot t^{\rm T}= 0$, etc.
where $t_\mu^{\rm T} = T_{\mu\nu}t_\nu$, etc., with $T_{\mu\nu}=\delta_{\mu\nu}- p_\mu p_\nu/p^2$.
%the standard idempotent transverse projection operator.
%
In order to simplify comparisons with Ref.\,\cite{Chang:2010hb}, we remark that the vectors in Fig.\,\ref{figvertex} must be mapped as follows: $q\to p_f$, $p_i\to k$, $k\to p$, and $t\to 2\ell$.  Then, denoting the Dirac-tensor basis used therein for the transverse vertex as $ \{ \hat T_\nu^i,i=1,\ldots,8\}$:
\begin{align}
\nonumber
\hat T_\nu^2 & = T_\nu^5\,,\;
\hat T_\nu^3 = T_\nu^3\,, \;
\hat T_\nu^4 = -T_\nu^8\,, \\
\hat T_\nu^5  & = - T_\nu^1 \,,\;
\hat T_\nu^7 = -2 T_\nu^2\,,\;
\hat T_\nu^8  = -2 T_\nu^4\,,
\end{align}
with $\hat T_\nu^{3,6}$ given by linear combinations of $T_\nu^{6,7}$, which are not needed herein.  Consequently, the coefficient functions in Ref.\,\cite{Chang:2010hb} are identified with ours thus:
\begin{align}
\nonumber
\hat F_1 & = \tau_3 \,,\;
\hat F_2 = \tau_5\,,\;
\hat F_4 = -\tau_8\,, \\
\hat F_5 & = -\tau_1\,,\;
\hat F_7 = -\tfrac{1}{2} \tau_2\,,\;
\hat F_8 = -\tfrac{1}{2} \tau_4\,.
\end{align}
\vspace*{-7ex}

%\smallskip

%\noindent\textbf{References}.

%%\bibliographystyle{../../zchanglei/zchanglei2/zPDA5/model1a-num-names}
%%\bibliography{../../../CollectedBiB}

\begin{thebibliography}{72}
\expandafter\ifx\csname natexlab\endcsname\relax\def\natexlab#1{#1}\fi
\providecommand{\bibinfo}[2]{#2}
\ifx\xfnm\relax \def\xfnm[#1]{\unskip,\space#1}\fi
%Type = Article
\bibitem[{Nambu and Jona-Lasinio(1961)}]{Nambu:1961tp}
\bibinfo{author}{Y.~Nambu}, \bibinfo{author}{G.~Jona-Lasinio},
  \bibinfo{journal}{Phys. Rev.} \bibinfo{volume}{122} (\bibinfo{year}{1961})
  \bibinfo{pages}{345--358}.
%Type = Article
\bibitem[{Lane(1974)}]{Lane:1974he}
\bibinfo{author}{K.~D. Lane}, \bibinfo{journal}{Phys. Rev. D}
  \bibinfo{volume}{10} (\bibinfo{year}{1974}) \bibinfo{pages}{2605}.
%Type = Article
\bibitem[{Politzer(1976)}]{Politzer:1976tv}
\bibinfo{author}{H.~D. Politzer}, \bibinfo{journal}{Nucl. Phys. B}
  \bibinfo{volume}{117} (\bibinfo{year}{1976}) \bibinfo{pages}{397}.
%Type = Article
\bibitem[{Bashir et~al.(2008)Bashir, Raya, Clo{\"e}t, and
  Roberts}]{Bashir:2008fk}
\bibinfo{author}{A.~Bashir}, \bibinfo{author}{A.~Raya}, \bibinfo{author}{I.~C.
  Clo{\"e}t}, \bibinfo{author}{C.~D. Roberts}, \bibinfo{journal}{Phys.\ Rev.\
  C} \bibinfo{volume}{78} (\bibinfo{year}{2008}) \bibinfo{pages}{055201}.
%Type = Article
\bibitem[{Bashir et~al.(2009)Bashir, Raya, S{\'a}nchez-Madrigal, and
  Roberts}]{Bashir:2009fv}
\bibinfo{author}{A.~Bashir}, \bibinfo{author}{A.~Raya},
  \bibinfo{author}{S.~S{\'a}nchez-Madrigal}, \bibinfo{author}{C.~D. Roberts},
  \bibinfo{journal}{Few Body Syst.} \bibinfo{volume}{46} (\bibinfo{year}{2009})
  \bibinfo{pages}{229--237}.
%Type = Article
\bibitem[{Raya et~al.(2013)}]{Raya:2013inaS}
\bibinfo{author}{K.~Raya}, et~al., \bibinfo{journal}{Phys. Rev. D}
  \bibinfo{volume}{88} (\bibinfo{year}{2013}) \bibinfo{pages}{096003}.
%Type = Article
\bibitem[{Chang et~al.(2013)Chang, Clo{\"e}t, Cobos-Martinez, Roberts, Schmidt,
  and Tandy}]{Chang:2013pq}
\bibinfo{author}{L.~Chang}, \bibinfo{author}{I.~C. Clo{\"e}t},
  \bibinfo{author}{J.~J. Cobos-Martinez}, \bibinfo{author}{C.~D. Roberts},
  \bibinfo{author}{S.~M. Schmidt}, \bibinfo{author}{P.~C. Tandy},
  \bibinfo{journal}{Phys. Rev. Lett.} \bibinfo{volume}{110}
  (\bibinfo{year}{2013}) \bibinfo{pages}{132001}.
%Type = Article
\bibitem[{Maris and Roberts(1997)}]{Maris:1997tm}
\bibinfo{author}{P.~Maris}, \bibinfo{author}{C.~D. Roberts},
  \bibinfo{journal}{Phys. Rev. C} \bibinfo{volume}{56} (\bibinfo{year}{1997})
  \bibinfo{pages}{3369--3383}.
%Type = Article
\bibitem[{Qin et~al.(2011)Qin, Chang, Liu, Roberts, and Wilson}]{Qin:2011dd}
\bibinfo{author}{S.-X. Qin}, \bibinfo{author}{L.~Chang}, \bibinfo{author}{Y.-X.
  Liu}, \bibinfo{author}{C.~D. Roberts}, \bibinfo{author}{D.~J. Wilson},
  \bibinfo{journal}{Phys. Rev. C} \bibinfo{volume}{84} (\bibinfo{year}{2011})
  \bibinfo{pages}{042202(R)}.
%Type = Article
\bibitem[{Qin et~al.(2012)Qin, Chang, Liu, Roberts, and Wilson}]{Qin:2011xq}
\bibinfo{author}{S.-X. Qin}, \bibinfo{author}{L.~Chang}, \bibinfo{author}{Y.-X.
  Liu}, \bibinfo{author}{C.~D. Roberts}, \bibinfo{author}{D.~J. Wilson},
  \bibinfo{journal}{Phys. Rev. C} \bibinfo{volume}{85} (\bibinfo{year}{2012})
  \bibinfo{pages}{035202}.
%Type = Article
\bibitem[{Jain and Munczek(1993)}]{Jain:1993qh}
\bibinfo{author}{P.~Jain}, \bibinfo{author}{H.~J. Munczek},
  \bibinfo{journal}{Phys. Rev. D} \bibinfo{volume}{48} (\bibinfo{year}{1993})
  \bibinfo{pages}{5403--5411}.
%Type = Article
\bibitem[{Bloch(2002)}]{Bloch:2002eq}
\bibinfo{author}{J.~C.~R. Bloch}, \bibinfo{journal}{Phys. Rev.}
  \bibinfo{volume}{D66} (\bibinfo{year}{2002}) \bibinfo{pages}{034032}.
%Type = Article
\bibitem[{Aguilar et~al.(2008)Aguilar, Binosi, and
  Papavassiliou}]{Aguilar:2008xm}
\bibinfo{author}{A.~Aguilar}, \bibinfo{author}{D.~Binosi},
  \bibinfo{author}{J.~Papavassiliou}, \bibinfo{journal}{Phys. Rev. D}
  \bibinfo{volume}{78} (\bibinfo{year}{2008}) \bibinfo{pages}{025010}.
%Type = Article
\bibitem[{Boucaud et~al.(2012)Boucaud, Leroy, Le-Yaouanc, Micheli, Pene, and
  Rodr{\'i}guez-Quintero}]{Boucaud:2011ug}
\bibinfo{author}{P.~Boucaud}, \bibinfo{author}{J.~P. Leroy},
  \bibinfo{author}{A.~Le-Yaouanc}, \bibinfo{author}{J.~Micheli},
  \bibinfo{author}{O.~Pene}, \bibinfo{author}{J.~Rodr{\'i}guez-Quintero},
  \bibinfo{journal}{Few Body Syst.} \bibinfo{volume}{53} (\bibinfo{year}{2012})
  \bibinfo{pages}{387--436}.
%Type = Article
\bibitem[{Aguilar et~al.(2016)Aguilar, Binosi, and
  Papavassiliou}]{Aguilar:2015bud}
\bibinfo{author}{A.~C. Aguilar}, \bibinfo{author}{D.~Binosi},
  \bibinfo{author}{J.~Papavassiliou}, \bibinfo{journal}{Front. Phys. China}
  \bibinfo{volume}{11} (\bibinfo{year}{2016}) \bibinfo{pages}{111203}.
%Type = Article
\bibitem[{Binosi et~al.(2015)Binosi, Chang, Papavassiliou, and
  Roberts}]{Binosi:2014aea}
\bibinfo{author}{D.~Binosi}, \bibinfo{author}{L.~Chang},
  \bibinfo{author}{J.~Papavassiliou}, \bibinfo{author}{C.~D. Roberts},
  \bibinfo{journal}{Phys. Lett. B} \bibinfo{volume}{742} (\bibinfo{year}{2015})
  \bibinfo{pages}{183--188}.
%Type = Article
\bibitem[{Cornwall(1982)}]{Cornwall:1981zr}
\bibinfo{author}{J.~M. Cornwall}, \bibinfo{journal}{Phys. Rev. D}
  \bibinfo{volume}{26} (\bibinfo{year}{1982}) \bibinfo{pages}{1453}.
%Type = Article
\bibitem[{Cornwall and Papavassiliou(1989)}]{Cornwall:1989gv}
\bibinfo{author}{J.~M. Cornwall}, \bibinfo{author}{J.~Papavassiliou},
  \bibinfo{journal}{Phys. Rev. D} \bibinfo{volume}{40} (\bibinfo{year}{1989})
  \bibinfo{pages}{3474}.
%Type = Article
\bibitem[{Binosi and Papavassiliou(2002)}]{Binosi:2002ez}
\bibinfo{author}{D.~Binosi}, \bibinfo{author}{J.~Papavassiliou},
  \bibinfo{journal}{Phys. Rev. D} \bibinfo{volume}{66} (\bibinfo{year}{2002})
  \bibinfo{pages}{025024}.
%Type = Article
\bibitem[{Binosi and Papavassiliou(2009)}]{Binosi:2009qm}
\bibinfo{author}{D.~Binosi}, \bibinfo{author}{J.~Papavassiliou},
  \bibinfo{journal}{Phys. Rept.} \bibinfo{volume}{479} (\bibinfo{year}{2009})
  \bibinfo{pages}{1--152}.
%Type = Article
\bibitem[{Aguilar et~al.(2009)Aguilar, Binosi, Papavassiliou, and
  Rodr{\'i}guez-Quintero}]{Aguilar:2009nf}
\bibinfo{author}{A.~Aguilar}, \bibinfo{author}{D.~Binosi},
  \bibinfo{author}{J.~Papavassiliou},
  \bibinfo{author}{J.~Rodr{\'i}guez-Quintero}, \bibinfo{journal}{Phys. Rev. D}
  \bibinfo{volume}{80} (\bibinfo{year}{2009}) \bibinfo{pages}{085018}.
%Type = Article
\bibitem[{Ball and Chiu(1980)}]{Ball:1980ay}
\bibinfo{author}{J.~S. Ball}, \bibinfo{author}{T.-W. Chiu},
  \bibinfo{journal}{Phys.\ Rev.\ D} \bibinfo{volume}{22} (\bibinfo{year}{1980})
  \bibinfo{pages}{2542--2549}.
%Type = Article
\bibitem[{Williams et~al.(1991)Williams, Krein, and Roberts}]{Williams:1989tv}
\bibinfo{author}{A.~G. Williams}, \bibinfo{author}{G.~Krein},
  \bibinfo{author}{C.~D. Roberts}, \bibinfo{journal}{Annals Phys.}
  \bibinfo{volume}{210} (\bibinfo{year}{1991}) \bibinfo{pages}{464--485}.
%Type = Article
\bibitem[{Curtis and Pennington(1990)}]{Curtis:1990zs}
\bibinfo{author}{D.~C. Curtis}, \bibinfo{author}{M.~R. Pennington},
  \bibinfo{journal}{Phys.\ Rev.\ D} \bibinfo{volume}{42} (\bibinfo{year}{1990})
  \bibinfo{pages}{4165--4169}.
%Type = Article
\bibitem[{Burden and Roberts(1991)}]{Burden:1990mg}
\bibinfo{author}{C.~J. Burden}, \bibinfo{author}{C.~D. Roberts},
  \bibinfo{journal}{Phys. Rev. D} \bibinfo{volume}{44} (\bibinfo{year}{1991})
  \bibinfo{pages}{540--550}.
%Type = Article
\bibitem[{Dong et~al.(1994)Dong, Munczek, and Roberts}]{Dong:1994jr}
\bibinfo{author}{Z.-H. Dong}, \bibinfo{author}{H.~J. Munczek},
  \bibinfo{author}{C.~D. Roberts}, \bibinfo{journal}{Phys. Lett.}
  \bibinfo{volume}{B333} (\bibinfo{year}{1994}) \bibinfo{pages}{536--544}.
%Type = Article
\bibitem[{Bashir and Pennington(1994)}]{Bashir:1994az}
\bibinfo{author}{A.~Bashir}, \bibinfo{author}{M.~Pennington},
  \bibinfo{journal}{Phys. Rev. D} \bibinfo{volume}{50} (\bibinfo{year}{1994})
  \bibinfo{pages}{7679--7689}.
%Type = Article
\bibitem[{Munczek(1995)}]{Munczek:1994zz}
\bibinfo{author}{H.~J. Munczek}, \bibinfo{journal}{Phys. Rev. D}
  \bibinfo{volume}{52} (\bibinfo{year}{1995}) \bibinfo{pages}{4736--4740}.
%Type = Article
\bibitem[{Stingl(1996)}]{Stingl:1994nk}
\bibinfo{author}{M.~Stingl}, \bibinfo{journal}{Z. Phys. A}
  \bibinfo{volume}{353} (\bibinfo{year}{1996}) \bibinfo{pages}{423--445}.
%Type = Article
\bibitem[{Bender et~al.(1996)Bender, Roberts, and von Smekal}]{Bender:1996bb}
\bibinfo{author}{A.~Bender}, \bibinfo{author}{C.~D. Roberts},
  \bibinfo{author}{L.~von Smekal}, \bibinfo{journal}{Phys. Lett. B}
  \bibinfo{volume}{380} (\bibinfo{year}{1996}) \bibinfo{pages}{7--12}.
%Type = Article
\bibitem[{Hawes et~al.(1998)Hawes, Maris, and Roberts}]{Hawes:1998cw}
\bibinfo{author}{F.~T. Hawes}, \bibinfo{author}{P.~Maris},
  \bibinfo{author}{C.~D. Roberts}, \bibinfo{journal}{Phys. Lett. B}
  \bibinfo{volume}{440} (\bibinfo{year}{1998}) \bibinfo{pages}{353--358}.
%Type = Article
\bibitem[{Bhagwat et~al.(2004)Bhagwat, H{\"o}ll, Krassnigg, Roberts, and
  Tandy}]{Bhagwat:2004hn}
\bibinfo{author}{M.~S. Bhagwat}, \bibinfo{author}{A.~H{\"o}ll},
  \bibinfo{author}{A.~Krassnigg}, \bibinfo{author}{C.~D. Roberts},
  \bibinfo{author}{P.~C. Tandy}, \bibinfo{journal}{Phys. Rev. C}
  \bibinfo{volume}{70} (\bibinfo{year}{2004}) \bibinfo{pages}{035205}.
%Type = Article
\bibitem[{Bhagwat and Tandy(2004)}]{Bhagwat:2004kj}
\bibinfo{author}{M.~S. Bhagwat}, \bibinfo{author}{P.~C. Tandy},
  \bibinfo{journal}{Phys. Rev. D} \bibinfo{volume}{70} (\bibinfo{year}{2004})
  \bibinfo{pages}{094039}.
%Type = Article
\bibitem[{Chang and Roberts(2009)}]{Chang:2009zb}
\bibinfo{author}{L.~Chang}, \bibinfo{author}{C.~D. Roberts},
  \bibinfo{journal}{Phys. Rev. Lett.} \bibinfo{volume}{103}
  (\bibinfo{year}{2009}) \bibinfo{pages}{081601}.
%Type = Article
\bibitem[{K{\i}z{\i}lers{\"u} and Pennington(2009)}]{Kizilersu:2009kg}
\bibinfo{author}{A.~K{\i}z{\i}lers{\"u}}, \bibinfo{author}{M.~R. Pennington},
  \bibinfo{journal}{Phys.\ Rev.\ D} \bibinfo{volume}{79} (\bibinfo{year}{2009})
  \bibinfo{pages}{125020}.
%Type = Article
\bibitem[{Chang et~al.(2011)Chang, Liu, and Roberts}]{Chang:2010hb}
\bibinfo{author}{L.~Chang}, \bibinfo{author}{Y.-X. Liu}, \bibinfo{author}{C.~D.
  Roberts}, \bibinfo{journal}{Phys. Rev. Lett.} \bibinfo{volume}{106}
  (\bibinfo{year}{2011}) \bibinfo{pages}{072001}.
%Type = Article
\bibitem[{Aguilar and Papavassiliou(2011)}]{Aguilar:2010cn}
\bibinfo{author}{A.~C. Aguilar}, \bibinfo{author}{J.~Papavassiliou},
  \bibinfo{journal}{Phys. Rev. D} \bibinfo{volume}{83} (\bibinfo{year}{2011})
  \bibinfo{pages}{014013}.
%Type = Article
\bibitem[{Qin et~al.(2013)Qin, Chang, Liu, Roberts, and Schmidt}]{Qin:2013mta}
\bibinfo{author}{S.-X. Qin}, \bibinfo{author}{L.~Chang}, \bibinfo{author}{Y.-X.
  Liu}, \bibinfo{author}{C.~D. Roberts}, \bibinfo{author}{S.~M. Schmidt},
  \bibinfo{journal}{Phys.\ Lett.\ B} \bibinfo{volume}{722}
  (\bibinfo{year}{2013}) \bibinfo{pages}{384--388}.
%Type = Article
\bibitem[{Aguilar et~al.(2014)Aguilar, Binosi, Iba{\~n}ez, and
  Papavassiliou}]{Aguilar:2014lha}
\bibinfo{author}{A.~C. Aguilar}, \bibinfo{author}{D.~Binosi},
  \bibinfo{author}{D.~Iba{\~n}ez}, \bibinfo{author}{J.~Papavassiliou},
  \bibinfo{journal}{Phys. Rev. D} \bibinfo{volume}{90} (\bibinfo{year}{2014})
  \bibinfo{pages}{065027}.
%Type = Article
\bibitem[{K{\i}z{\i}lers{\"u} et~al.(2015)K{\i}z{\i}lers{\"u}, Sizer,
  Pennington, Williams, and Williams}]{Kizilersu:2014ela}
\bibinfo{author}{A.~K{\i}z{\i}lers{\"u}}, \bibinfo{author}{T.~Sizer},
  \bibinfo{author}{M.~R. Pennington}, \bibinfo{author}{A.~G. Williams},
  \bibinfo{author}{R.~Williams}, \bibinfo{journal}{Phys. Rev. D}
  \bibinfo{volume}{91} (\bibinfo{year}{2015}) \bibinfo{pages}{065015}.
%Type = Article
\bibitem[{Mitter et~al.(2015)Mitter, Pawlowski, and
  Strodthoff}]{Mitter:2014wpa}
\bibinfo{author}{M.~Mitter}, \bibinfo{author}{J.~M. Pawlowski},
  \bibinfo{author}{N.~Strodthoff}, \bibinfo{journal}{Phys. Rev. D}
  \bibinfo{volume}{91} (\bibinfo{year}{2015}) \bibinfo{pages}{054035}.
%Type = Article
\bibitem[{Binosi et~al.(2016)Binosi, Chang, Papavassiliou, Qin, and
  Roberts}]{Binosi:2016rxz}
\bibinfo{author}{D.~Binosi}, \bibinfo{author}{L.~Chang},
  \bibinfo{author}{J.~Papavassiliou}, \bibinfo{author}{S.-X. Qin},
  \bibinfo{author}{C.~D. Roberts}, \bibinfo{journal}{Phys. Rev. D}
  \bibinfo{volume}{93} (\bibinfo{year}{2016}) \bibinfo{pages}{096010}.
%Type = Article
\bibitem[{Chang and Roberts(2012)}]{Chang:2011ei}
\bibinfo{author}{L.~Chang}, \bibinfo{author}{C.~D. Roberts},
  \bibinfo{journal}{Phys. Rev. C} \bibinfo{volume}{85} (\bibinfo{year}{2012})
  \bibinfo{pages}{052201(R)}.
%Type = Article
\bibitem[{Sanchis-Alepuz and Williams(2015)}]{Sanchis-Alepuz:2015qra}
\bibinfo{author}{H.~Sanchis-Alepuz}, \bibinfo{author}{R.~Williams},
  \bibinfo{journal}{Phys. Lett. B} \bibinfo{volume}{749} (\bibinfo{year}{2015})
  \bibinfo{pages}{592--596}.
%Type = Article
\bibitem[{Qin(2016)}]{Qin:2016fbu}
\bibinfo{author}{S.-X. Qin}, \bibinfo{journal}{EPJ Web Conf.}
  \bibinfo{volume}{113} (\bibinfo{year}{2016}) \bibinfo{pages}{05024}.
%Type = Article
\bibitem[{Takahashi(1985)}]{Takahashi:1985yz}
\bibinfo{author}{Y.~Takahashi}  (\bibinfo{year}{1985}).
  \bibinfo{note}{{\emph{Canonical quantization and generalized Ward relations:
  Foundation of nonperturbative approach}, Print-85-0421 (Alberta)}}.
%Type = Article
\bibitem[{He et~al.(2000)He, Khanna, and Takahashi}]{He:2000we}
\bibinfo{author}{H.-X. He}, \bibinfo{author}{F.~Khanna},
  \bibinfo{author}{Y.~Takahashi}, \bibinfo{journal}{Phys. Lett.}
  \bibinfo{volume}{B480} (\bibinfo{year}{2000}) \bibinfo{pages}{222--228}.
%Type = Article
\bibitem[{Pennington and Williams(2006)}]{Pennington:2005mw}
\bibinfo{author}{M.~Pennington}, \bibinfo{author}{R.~Williams},
  \bibinfo{journal}{J. Phys. G} \bibinfo{volume}{32} (\bibinfo{year}{2006})
  \bibinfo{pages}{2219--2234}.
%Type = Article
\bibitem[{He(2009)}]{He:2009sj}
\bibinfo{author}{H.-X. He}, \bibinfo{journal}{Phys. Rev. D}
  \bibinfo{volume}{80} (\bibinfo{year}{2009}) \bibinfo{pages}{016004}.
%Type = Article
\bibitem[{Skullerud and Kizilersu(2002)}]{Skullerud:2002ge}
\bibinfo{author}{J.~Skullerud}, \bibinfo{author}{A.~Kizilersu},
  \bibinfo{journal}{JHEP} \bibinfo{volume}{09} (\bibinfo{year}{2002})
  \bibinfo{pages}{013}.
%Type = Article
\bibitem[{Lin(2006)}]{Lin:2005zd}
\bibinfo{author}{H.-W. Lin}, \bibinfo{journal}{Phys. Rev. D}
  \bibinfo{volume}{73} (\bibinfo{year}{2006}) \bibinfo{pages}{094511}.
%Type = Article
\bibitem[{K{\i}z{\i}lers{\"u} et~al.(2007)K{\i}z{\i}lers{\"u}, Leinweber,
  Skullerud, and Williams}]{Kizilersu:2006et}
\bibinfo{author}{A.~K{\i}z{\i}lers{\"u}}, \bibinfo{author}{D.~B. Leinweber},
  \bibinfo{author}{J.-I. Skullerud}, \bibinfo{author}{A.~G. Williams},
  \bibinfo{journal}{Eur. Phys. J.} \bibinfo{volume}{50} (\bibinfo{year}{2007})
  \bibinfo{pages}{871--875}.
%Type = Article
\bibitem[{Rojas et~al.(2013)Rojas, de~Melo, El-Bennich, Oliveira, and
  Frederico}]{Rojas:2013tza}
\bibinfo{author}{E.~Rojas}, \bibinfo{author}{J.~P. B.~C. de~Melo},
  \bibinfo{author}{B.~El-Bennich}, \bibinfo{author}{O.~Oliveira},
  \bibinfo{author}{T.~Frederico}, \bibinfo{journal}{JHEP} \bibinfo{volume}{10}
  (\bibinfo{year}{2013}) \bibinfo{pages}{193}.
%Type = Article
\bibitem[{Bashir et~al.(2012)Bashir, Berm{\'u}dez, Chang, and
  Roberts}]{Bashir:2011dp}
\bibinfo{author}{A.~Bashir}, \bibinfo{author}{R.~Berm{\'u}dez},
  \bibinfo{author}{L.~Chang}, \bibinfo{author}{C.~D. Roberts},
  \bibinfo{journal}{Phys.\ Rev.\ C} \bibinfo{volume}{85} (\bibinfo{year}{2012})
  \bibinfo{pages}{045205}.
%Type = Article
\bibitem[{Gasser and Leutwyler(1984)}]{Gasser:1983yg}
\bibinfo{author}{J.~Gasser}, \bibinfo{author}{H.~Leutwyler},
  \bibinfo{journal}{Annals Phys.} \bibinfo{volume}{158} (\bibinfo{year}{1984})
  \bibinfo{pages}{142}.
%Type = Article
\bibitem[{Pagels and Stokar(1979)}]{Pagels:1979hd}
\bibinfo{author}{H.~Pagels}, \bibinfo{author}{S.~Stokar},
  \bibinfo{journal}{Phys. Rev.} \bibinfo{volume}{D20} (\bibinfo{year}{1979})
  \bibinfo{pages}{2947}.
%Type = Article
\bibitem[{Cahill and Roberts(1985)}]{Cahill:1985mh}
\bibinfo{author}{R.~T. Cahill}, \bibinfo{author}{C.~D. Roberts},
  \bibinfo{journal}{Phys. Rev.} \bibinfo{volume}{D32} (\bibinfo{year}{1985})
  \bibinfo{pages}{2419}.
%Type = Article
\bibitem[{Maris et~al.(1998)Maris, Roberts, and Tandy}]{Maris:1997hd}
\bibinfo{author}{P.~Maris}, \bibinfo{author}{C.~D. Roberts},
  \bibinfo{author}{P.~C. Tandy}, \bibinfo{journal}{Phys. Lett. B}
  \bibinfo{volume}{420} (\bibinfo{year}{1998}) \bibinfo{pages}{267--273}.
%Type = Article
\bibitem[{Clo{\"e}t et~al.(2013)Clo{\"e}t, Roberts, and Thomas}]{Cloet:2013gva}
\bibinfo{author}{I.~C. Clo{\"e}t}, \bibinfo{author}{C.~D. Roberts},
  \bibinfo{author}{A.~W. Thomas}, \bibinfo{journal}{Phys. Rev. Lett.}
  \bibinfo{volume}{111} (\bibinfo{year}{2013}) \bibinfo{pages}{101803}.
%Type = Article
\bibitem[{Davydychev et~al.(2001)Davydychev, Osland, and
  Saks}]{Davydychev:2000rt}
\bibinfo{author}{A.~I. Davydychev}, \bibinfo{author}{P.~Osland},
  \bibinfo{author}{L.~Saks}, \bibinfo{journal}{Phys. Rev. D}
  \bibinfo{volume}{63} (\bibinfo{year}{2001}) \bibinfo{pages}{014022}.
%Type = Article
\bibitem[{Maris and Tandy(1999)}]{Maris:1999nt}
\bibinfo{author}{P.~Maris}, \bibinfo{author}{P.~C. Tandy},
  \bibinfo{journal}{Phys. Rev. C} \bibinfo{volume}{60} (\bibinfo{year}{1999})
  \bibinfo{pages}{055214}.
%Type = Article
\bibitem[{Nguyen et~al.(2011)Nguyen, Souchlas, and Tandy}]{Nguyen:2010yh}
\bibinfo{author}{T.~Nguyen}, \bibinfo{author}{N.~A. Souchlas},
  \bibinfo{author}{P.~C. Tandy}, \bibinfo{journal}{AIP Conf. Proc.}
  \bibinfo{volume}{1361} (\bibinfo{year}{2011}) \bibinfo{pages}{142--151}.
%Type = Article
\bibitem[{Clo{\"e}t et~al.(2013)Clo{\"e}t, Chang, Roberts, Schmidt, and
  Tandy}]{Cloet:2013tta}
\bibinfo{author}{I.~C. Clo{\"e}t}, \bibinfo{author}{L.~Chang},
  \bibinfo{author}{C.~D. Roberts}, \bibinfo{author}{S.~M. Schmidt},
  \bibinfo{author}{P.~C. Tandy}, \bibinfo{journal}{Phys. Rev. Lett.}
  \bibinfo{volume}{111} (\bibinfo{year}{2013}) \bibinfo{pages}{092001}.
%Type = Article
\bibitem[{Ding et~al.(2016)Ding, Gao, Chang, Liu, and Roberts}]{Ding:2015rkn}
\bibinfo{author}{M.~Ding}, \bibinfo{author}{F.~Gao},
  \bibinfo{author}{L.~Chang}, \bibinfo{author}{Y.-X. Liu},
  \bibinfo{author}{C.~D. Roberts}, \bibinfo{journal}{Phys. Lett. B}
  \bibinfo{volume}{753} (\bibinfo{year}{2016}) \bibinfo{pages}{330--335}.
%Type = Article
\bibitem[{Horn and Roberts(2016)}]{Horn:2016rip}
\bibinfo{author}{T.~Horn}, \bibinfo{author}{C.~D. Roberts},
  \bibinfo{journal}{J. Phys. G.} \bibinfo{volume}{43} (\bibinfo{year}{2016})
  \bibinfo{pages}{073001/1--47}.
%Type = Article
\bibitem[{Krassnigg(2009)}]{Krassnigg:2009zh}
\bibinfo{author}{A.~Krassnigg}, \bibinfo{journal}{Phys. Rev. D}
  \bibinfo{volume}{80} (\bibinfo{year}{2009}) \bibinfo{pages}{114010}.
%Type = Article
\bibitem[{Krassnigg and Blank(2011)}]{Krassnigg:2010mh}
\bibinfo{author}{A.~Krassnigg}, \bibinfo{author}{M.~Blank},
  \bibinfo{journal}{Phys. Rev. D} \bibinfo{volume}{83} (\bibinfo{year}{2011})
  \bibinfo{pages}{096006}.
%Type = Article
\bibitem[{Roberts et~al.(2011)Roberts, Chang, Clo{\"e}t, and
  Roberts}]{Roberts:2011cf}
\bibinfo{author}{H.~L.~L. Roberts}, \bibinfo{author}{L.~Chang},
  \bibinfo{author}{I.~C. Clo{\"e}t}, \bibinfo{author}{C.~D. Roberts},
  \bibinfo{journal}{Few Body Syst.} \bibinfo{volume}{51} (\bibinfo{year}{2011})
  \bibinfo{pages}{1--25}.
%Type = Article
\bibitem[{Chen et~al.(2012)Chen, Chang, Roberts, Wan, and Wilson}]{Chen:2012qr}
\bibinfo{author}{C.~Chen}, \bibinfo{author}{L.~Chang}, \bibinfo{author}{C.~D.
  Roberts}, \bibinfo{author}{S.-L. Wan}, \bibinfo{author}{D.~J. Wilson},
  \bibinfo{journal}{Few Body Syst.} \bibinfo{volume}{53} (\bibinfo{year}{2012})
  \bibinfo{pages}{293--326}.
%Type = Article
\bibitem[{Rojas et~al.(2014)Rojas, El-Bennich, and de~Melo}]{Rojas:2014aka}
\bibinfo{author}{E.~Rojas}, \bibinfo{author}{B.~El-Bennich},
  \bibinfo{author}{J.~P. B.~C. de~Melo}, \bibinfo{journal}{Phys. Rev. D}
  \bibinfo{volume}{90} (\bibinfo{year}{2014}) \bibinfo{pages}{074025}.
%Type = Article
\bibitem[{G\'{o}mez-Rocha et~al.(2015)G\'{o}mez-Rocha, Hilger, and
  Krassnigg}]{Gomez-Rocha:2015qga}
\bibinfo{author}{M.~G\'{o}mez-Rocha}, \bibinfo{author}{T.~Hilger},
  \bibinfo{author}{A.~Krassnigg}, \bibinfo{journal}{Phys. Rev. D}
  \bibinfo{volume}{92} (\bibinfo{year}{2015}) \bibinfo{pages}{054030}.
%Type = Article
\bibitem[{Eichmann et~al.(p ph)Eichmann, Sanchis-Alepuz, Williams, Alkofer, and
  Fischer}]{Eichmann:2016yit}
\bibinfo{author}{G.~Eichmann}, \bibinfo{author}{H.~Sanchis-Alepuz},
  \bibinfo{author}{R.~Williams}, \bibinfo{author}{R.~Alkofer},
  \bibinfo{author}{C.~S. Fischer}  (\bibinfo{year}{arXiv:1606.09602 [hep-ph]}).
  \bibinfo{note}{{\emph{Baryons as relativistic three-quark bound states}}}.

\end{thebibliography}

\end{document}